\documentclass[12pt, draftclsnofoot, onecolumn]{IEEEtran}
\usepackage{blindtext}
\usepackage{graphicx}
\usepackage{bigints}
\usepackage{caption,float}
\newlength\figureheight 
\newlength\figurewidth 
\usepackage{cite}
\usepackage{url}
\usepackage{blindtext, graphicx}
\usepackage{amsmath,amssymb,mathtools,bm,etoolbox}

\usepackage{cuted}
\usepackage{multicol,lipsum}
\usepackage{multirow}
\usepackage{breqn}
\newcommand\ddfrac[2]{\frac{\displaystyle #1}{\displaystyle #2}}

\usepackage{boxhandler}

\newcommand{\e}[1]{{\mathbb E}\left[ #1 \right]}
\usepackage{stackengine}
\def\delequal{\mathrel{\ensurestackMath{\stackon[1pt]{=}{\scriptstyle\Delta}}}}
\usepackage{suffix}

\DeclarePairedDelimiterX\MeijerM[3]{\lparen}{\rparen}%
{\begin{smallmatrix}#1 \\ #2\end{smallmatrix}\delimsize\vert\,#3}

\newcommand\MeijerG[8][]{%
  G^{\,#2,#3}_{#4,#5}\MeijerM[#1]{#6}{#7}{#8}}

\WithSuffix\newcommand\MeijerG*[7]{%
  G^{\,#1,#2}_{#3,#4}\MeijerM*{#5}{#6}{#7}}
\usepackage{ellipsis}
\usepackage{tkz-euclide}
\usepackage{tikz}
\usepackage{tikz-3dplot}
\usepackage[utf8]{inputenc}
\usepackage{pgfplots} 
\usepackage{pgfgantt}
\usepackage{pdflscape}
\pgfplotsset{compat=newest} 
\pgfplotsset{plot coordinates/math parser=false}
\pgfplotsset{every  tick/.style={black,},ylabel style={font=\tiny},xlabel style={font=\tiny},tick label style={font=\tiny},legend style= {font=\scriptsize},
minor x tick num=1,minor y tick num=1,xminorticks=true,yminorticks=true,}
  \newlength\fheight
\newlength\fwidth

\usetikzlibrary{plotmarks}

\begin{document}

\title{Asymmetric RF/FSO Relaying with HPA non-Linearities and Feedback Delay Constraints}

\author{Elyes~Balti,~\IEEEmembership{Student Member,~IEEE,}
        and~Brian~K.~Johnson,~\IEEEmembership{Fellow,~IEEE}}%

\maketitle

\begin{abstract}
In this work, we investigate the performance of a dual-hop multiple relays system consisting of mixed Radio-Frequency (RF)/Free Space Optical (FSO) channels. The RF channels are subject to Rayleigh fading while the optical links experience the Double Generalized Gamma including atmospheric turbulence, path loss and the misalignment  between the transmitter and the receiver aperture (also known as the pointing error). The FSO model also takes into account the receiver detection technique which could be either heterodyne or intensity modulation and direct detection. Partial Relay Selection with outdated Channel State Information is assumed based on the RF channels to select a relay and we also consider fixed and variable Amplify-and-Forward relaying schemes. In addition, we assume that the relays are affected by the high power amplifier non-linearities and herein we discuss two power amplifiers called Soft Envelope Limiter and Traveling Wave Tube Amplifier. Furthermore, novel closed-forms and tight upper bounds of the outage probability, the bit error probability, and the ergodic capacity are derived. Capitalizing on these performance, we derive the high SNR asymptotic to get engineering insights about the system gains such as the diversity and the coding gains. Finally, the mathematical expressions are validated using the Monte Carlo simulation.  
\end{abstract}


\IEEEpeerreviewmaketitle

\section{Introduction}
\IEEEPARstart{W}{ith} the rapid increase of the internet base and mobile stations and the extremely high demand for bandwidth, the Radio Frequency (RF) cellular systems have reached a saturation level owing to the limited spectrum  and expensive access licence. Although, many research attempts in cognitive radio allow parallel utilization of the bandwidth between the primary and secondary users, the last ones still suffer from the spectrum drought since they are always benefiting from some spectrum holes left by the primary users. Moreover, the backhaul network cannot support the big data flow even for the licenced primary ones. Recent attempts have proposed the usage of optical fibers (OF) as a solution for the backhaul network congestion. However, for ultra dense cellular networks, a large number of OF are needed given that these cable installations are very costly and the space installation of such cables to serve large number of cells/users are limited and even restricted in some areas. To support such network densification, Free Space Optical (FSO) technology has been recently proposed as an alternative or complementary solution to both RF and OF due to its flexibility, free spectrum access licence, power efficiency, cost effectiveness, no installation restriction and most importantly it is a way to densify the cellular networks with limited congestion and delays \cite{5,8}. Due to these advantages, FSO is seen as the corner stone of the-fifth generation (5G) since it is predicted to achieve 25 times the average cell throughput, 10 times the spectral/energy efficiency, 1000 times the system capacity and from 10 to 100 times the data rate compared to the LTE or the fourth-generation (4G) \cite{10,mmwave}. Besides, FSO systems employ a narrow laser beam which offers a high security level, immunity to electromagnetic interference and operating frequencies above 300 GHz. Because of these advantages, FSO technology has been considered as a possible solution for the last mile problem to bridge the bandwidth gap between the end-users and the OF backbone network. Moreover, the FSO technology has been also applied over the following applications such as enterprise/campus connectivity, video surveillance and monitoring, backhaul for cellular systems, redundant link and disaster recovery, security and broadcasting \cite{5}. 
\subsection{Motivation}
 FSO technology becomes a reliable and promising technique which has recently gained enormous interests especially in mixed RF/FSO systems. Previous work have proposed various channel models for the optical fading. In fact, Log-normal distribution is widely employed to statistically model the optical irradiance \cite{ln1} since it provides a good fit to the experimental data for weak turbulence. However, Log-normal model largely deviates from the experimental data as the atmospheric turbulence becomes more severe. To overcome this shortcoming, recent work have proposed the so-called Gamma-Gamma (G$^2$) \cite{60} as a model for the FSO fading since it provides a good fit to the experimental data for a wider range of the atmospheric turbulences compared to the Log-normal distribution. However, G$^2$ fails to provide a good fit with the experimental data especially at the tails. Since the calculation of the fade and the detection probability are essentially based on the tail of the probability density function (PDF), underestimation or overestimation of the tail region affects the performance analysis accuracy and certainly leads to erroneous results. To address this problem, Kashani \textit{et. al} \cite{dgg} introduced a new efficient optical fading model called Double Generalized Gamma which not only reflects a wide range of the atmospheric turbulences but also it provides a good fit to the experimental data particularly at the tail region.\\
As the optical signal propagates in free-space, it is susceptible not only to the atmospheric turbulences but also to the path loss and the pointing error as well. The path loss is basically depends on the link distance and the atmospheric attenuation which describes the weather conditions going from clear air, hazy, rainy and foggy. The work \cite{ln1,43} provide some typical values of the atmospheric attenuations describing the corresponding weather conditions. Moreover, the optical signal is also subject to the pointing error which can be described as the misalignment between the laser-emitting relay and the receiver photodetector. In fact, this misalignment is mainly caused by the building sway and seismic activities resulting in the pointing error that may arise severely as the relays and the receiver are located on tall buildings. The pointing error can be interpreted as an additional FSO fading that requires an accurate model to quantify its impact on the FSO signal. Uysal \textit{et. al} \cite{44} have proposed various models for the radial displacement of the pointing error assuming a Gaussian laser beam. The most general model proposed is called Beckmann pointing error model and there are various special cases derived from it. Previous work have assumed that the radial displacement can be modeled as Rician \cite{45}, Hoyt \cite{46}, NonZero-Mean and Zero-Mean Single-Sided Gaussian \cite{47} but the most prevalent one is Rayleigh \cite{48,int} for simplicity. \\
Furthemore, the optical signal could be detected following different schemes and the most widely used are the heterodyne and intensity-modulation and direct detection (IM/DD) \cite{2}. Although previous work have shown that the heterodyne configuration outperforms IM/DD, it is still hard to be implemented in the system. As a result, recent work have focused on IM/DD with on-off keying (OOK) due to its cost effective and easy implementation, however, this scheme requires an adaptive threshold for the demodulation \cite{2}. To address this shortcoming, the subcarrier intensity modulation (SIM) has been suggested as an alternative to IM/DD with OOK since this technique states that the RF signal is premodulated before the laser modulation \cite{50}.\\
It is true that the FSO contributes in densifying the number of users, the cellular networks still suffer from low signal coverage in some areas mainly located in forests and mountains where the optical signal cannot travel to such long distances and it is also heavily absorbed by the intermediate objects due to its high frequency. In an attempt to increase the coverage and the scalability of the network, one way is to implement the relays between the source ($S$) and the destination ($D$). Because of this advantage, cooperative relaying-assisted communication is considered as one of the key technologies for the next generation wireless communications because it plays an important role in improving the Quality of Service (QoS), reliability and coverage \cite{11}. The majority of the research attempts investigated mixed RF/FSO system considering various relaying schemes. The most prominent ones are Amplify-and-Forward (AF) \cite{aggregate,glob2017}, Decode-and-Forward (DF) \cite{17}, Quantize-and-Encode (QE) \cite{19}, and Quantize-and-Forward (QF) \cite{20}. Moreover, many research attempts have assumed systems employing either single or multiple relays. For the single relay system, there is only one way to forward the signal to the destination through the relay. However, for multiple relay systems, there are two possible ways either sending parallel transmissions when simultaneously activate all the relays or selecting one relay among the total set. In fact, there are many relay selection protocols such as opportunistic relay selection, partial relay selection \cite{21}, distributed switch and stay, max-select protocol and all active relaying \cite{24}. The latter is not well recommended since the receiver will suffer from the synchronization problem which occurs when using optical communications.\\ 
Although many contributions of the mixed RF/FSO system are presented and validated but these attempts considered ideal hardware without impairments. In fact, these impairments can be neglected for low rate systems but cannot be omitted in case of high rate systems especially when we introduce optics in order to improve the transfer rate. In practice, hardware always suffers from impairments, e.g., High Power Amplifier (HPA) non-linearities \cite{27}, phase noise \cite{28} and IQ imbalance \cite{30}. Given that the relays have low-cost, they are certainly of low quality and hence their tranceivers are more prone to impairments. Qi \textit{et. al} \cite{31} concluded that the impairments have deleterious effects on the system by limiting its performance in terms of outage probability, symbol error rate and channel capacity especially, in the high Signal-to-Noise Ratio (SNR) regime. In fact, previous work \cite{32} demonstrated that the impaired systems have a finite capacity limit at high SNR while there are floors for both the outage probability and the symbol error rate \cite{33}. Regarding the HPA non-linearities, this impairment is originated by the non-linear relaying amplification and as a result a non linear distortion is created and affects substantially the quality of the signal. In practice, there is a finite maximum output level for which any power amplifier can produce it and such saturation level is basically amplifier-dependent and varies to some extent but regardless of the amplifier model, this ceiling level is always bounded. In case when the power amplifier becomes unable to produce such power level, a signal distortion over the peak may arise and such phenomena is called clipping (clipping factor) of the power amplifier. In addition, the HPA model can be classified into two categories which are memoryless HPA and HPA with memory. The HPA is considered memoryless or frequency-independent if its frequency response characteristics are flat over the operating frequency range and in this case, the HPA is fully characterized by the two characteristics AM/AM (amplitude to amplitude conversion) and AM/PM (amplitude to phase conversion). On the other hand, the HPA is said to be with memory if its frequency responce characteristics are totally dependent on the frequency components or to the thermal phenomena \cite{42}. Such model can be classified as Hammerstein system that can be modeled by a series of a memoryless HPA and a linear filter. There are many types of this impairment that have been already covered in the literature but the most widely used are Soft Envelope Limiter (SEL), Traveling Wave Tube Amplifier (TWTA) and Solid State Power Amplifier (SSPA) or also called the Rapp model \cite{41}. The SEL is typically used to model a HPA with a perfect predistortion system while the TWTA has been primarily considered to model the non-linearities effect in OFDM system. However, the SSPA is characterized by a smoothness factor to control the switching between the saturation and the linear ranges. This model effectively discusses a linear characteristic for low magnitudes of the input signal and then it is limited by a definite constant saturated output. As the smoothness factor grows largely to infinity, this HPA model becomes the SEL impairments model. 
\subsection{Literature}
The existing work of the mixed RF/FSO systems cover various permutations of the system parameters. The authors in \cite{55,56} consider dual-hop hybrid RF/FSO system employing AF with fixed gain (FG). Particularly, Zedini \textit{et. al} in \cite{55} derive the outage probability, the bit error rate (BER) and the ergodic capacity assuming that the RF and FSO follows Nakagami-m and unified G$^2$, respectively. Besides, Al-Quwaiee \textit{et. al} in \cite{56} present the same performance as the aforementioned work but they assume that the RF and FSO channels experience Rayleigh and Double Generalized Gamma fading, respectively. On the other side, \cite{58,59} develop asymmetric dual-hop mixed RF/FSO systems with variable gain (VG). Ansari \textit{et. al} in \cite{58} derive novel closed-forms of the outage probability, BER and the average capacity where the RF and FSO links experience Rayleigh and unified G$^2$ while Yang \textit{et. al} in \cite{59} derive the same performance achieved by \cite{58} but they assume transmit diversity at the source and selection combining at the receiver. In addition the RF links are subject to Nakagami-m while the FSO fading is modeled by M{\'a}laga distribution. Further work \cite{60,61} assume mixed RF/FSO multiple relays systems with outdated CSI and they extend their work compared to the previous attempts by considering non-ideal hardware suffering from an aggregate model of hardware impairments. Although, the aforementioned work have considered many permutations of the system parameters, they did not consider more realistic and practical RF/FSO system including both the spatial diversity brought by the multiple relays and a particular model of the HPA non-linearities rather than assuming a general model of impairments. Hence, the contribution of this work is the objective of the next subsection.
\subsection{Contribution}
In this paper, we introduce two impairment models SEL and TWTA to the relays over a dual-hop mixed RF/FSO system with multiple relays. As a strategy to select the best candidate relay, we adopt the partial relay selection protocol with outdated channel state information (CSI) based on the partial knowledge of the first hop. In fact, the channels are generally time-varying and due to the slow feedback delay from the relays to the source, the CSI used for the relay selection is outdated and so the selected relay is not necessarily the best choice. Moreover, we consider AF for both Fixed Gain (FG) and Variable Gain (VG) relaying and we assume that the optical signal can be detected following either heterodyne or IM/DD while a subcarrier signal is used to modulate the intensity of an optical carrier (representing SIM technique). We also consider different types of modulation to get accurate insights into the study of the bit error probability under the conditions of the impairments.
To the best of our knowledge, this is the first work presenting a global analytical framework of mixed RF/FSO system with multiple relays suffering from various types of impairments. The contribution of this work are as follows:
\begin{itemize}
    \item Present a detailed description of the system architecture and the different models of impairments, we then take into account a macroscopic analysis and study the impact of the hardware impairments on the system performance.
    \item Specify the statistics of the RF and the optical channels in terms of the probability density function (PDF), the cumulative distribution function (CDF) and the high order moments.
    \item  After calculating the end-to-end Signal-to-Noise-plus-Distortion Ratio (SNDR) for both FG and VG relaying, we present the analytical formulations of the outage probability, the bit error probability, the ergodic capacity, the upper bounds and the asymptotic high SNR for SEL and TWTA and for various system parameters permutations such as the time correlation of the CSI, the atmospheric turbulence condition, the number of the relays, the rank of the selected relays, the path loss and the pointing error. Once the impacts of these parameters are quantified on the system performance, we can derive quantitative summaries and valuable engineering insights to draw meaningful conclusions and observations of the proposed system.
\end{itemize}
\subsection{Structure}
This paper is organized as follows: section II describes the system and the HPA non-linearities models. The system performance in terms of the outage probabilty, the bit error probability and the ergodic capacity analysis for FG and VG relaying are presented in section III and IV, respectively. Numerical results and their discussions are given in section V. The final section discusses the summary of this work.
\subsection{Notation}
For the sake of organization, we provide some useful notations to avoid the repetition. $f_{\rm h}(\cdot)$ and $F_{\rm h}(\cdot)$ denote the PDF and CDF of the random variable $h$, respectively. The Generalized Gamma distribution with parameters $\alpha, \beta$ and $\gamma$ is given by $\mathcal{GG} (\alpha,\beta,\gamma)$. In addition, the Gaussian distribution of parameter $\mu, \sigma^2$ is denoted by $\mathcal{N}(\mu,\sigma^2)$. The operator $\e{\cdot}$ stands for the expectation while Pr($\cdot$) denotes the probability measure. The symbol $\backsim$ stands for "distributed as".
\vspace*{-1cm}
\section{System and Channels Models}
\subsection{System Model}
\subsubsection*{\textbf{Relay Selection Protocol}}
Our system consists of $S$, $D$, and $N$ relays wirelessly linked to $S$ and $D$. As mentioned earlier, these relays amplify the incoming signal and then forward it to the destination. The amplification gain can be either FG or VG. FG relaying consists of amplifying the signal based on the average received SNR. However, VG relaying consists of amplifying the signal based on the received instantaneous SNR. To select the candidate relay of rank \textit{m}, we refer to the Partial Relay Selection (PRS) with outdated CSI to pick the best one based on the local feedbacks of the RF channels.
For a given communication, $S$ receives local feedback ($\gamma_{1(i)}$ for \textit{i = 1,\ldots N}) of the first hop obtained by the channel estimation from the \textit{N} relays and arranges them in an increasing order of amplitudes as follows: $\gamma_{1(1)}\leq\gamma_{1(2)}\leq \ldots \leq\gamma_{1(N)}$. The best scenario is to select the best relay \textit{(m = N)}. However, the best relay is not always available, so $S$ will pick the next best available relay. Thus PRS consists of selecting the \textit{m}-th worst or \textit{(N - m)}-th best relay $R_{(m)}$. Given that the local feedback coming from the relays to $S$ are very slow and the channels are very time-varying, the CSI that is used for the relay selection is not the same as the CSI used for the transmission. In this case, an outdated CSI must be considered instead of a perfect CSI. As a result, the current and outdated CSI are correlated with the correlation coefficient $\rho$ as follows
\begin{equation} \label{eq:1}
h_{1(m)} = \sqrt{\rho}~\hat{h}_{1(m)} + \sqrt{1-\rho}~\omega_{m}, 
\end{equation}
where $\omega_{m}$ is a random variable that follows the circularly complex Gaussian distribution with the same variance of the channel gain $h_{1(m)}$. The correlation coefficient $\rho$ is given by the Jakes' autocorrelation model \cite{jakes} as follows
\begin{equation}\label{eq:2}
\rho = J_0(2\pi f_{d} T_d), 
\end{equation}
where $J_0(\cdot)$ is the zeroth order Bessel function of the first kind, $T_d$ is the time delay between the current and the delayed CSI versions and $f_d$ is the maximum Doppler frequency of the channels.
\begin{figure}[H]
\centering
\setlength\fheight{6cm}
\setlength\fwidth{12cm}
\input{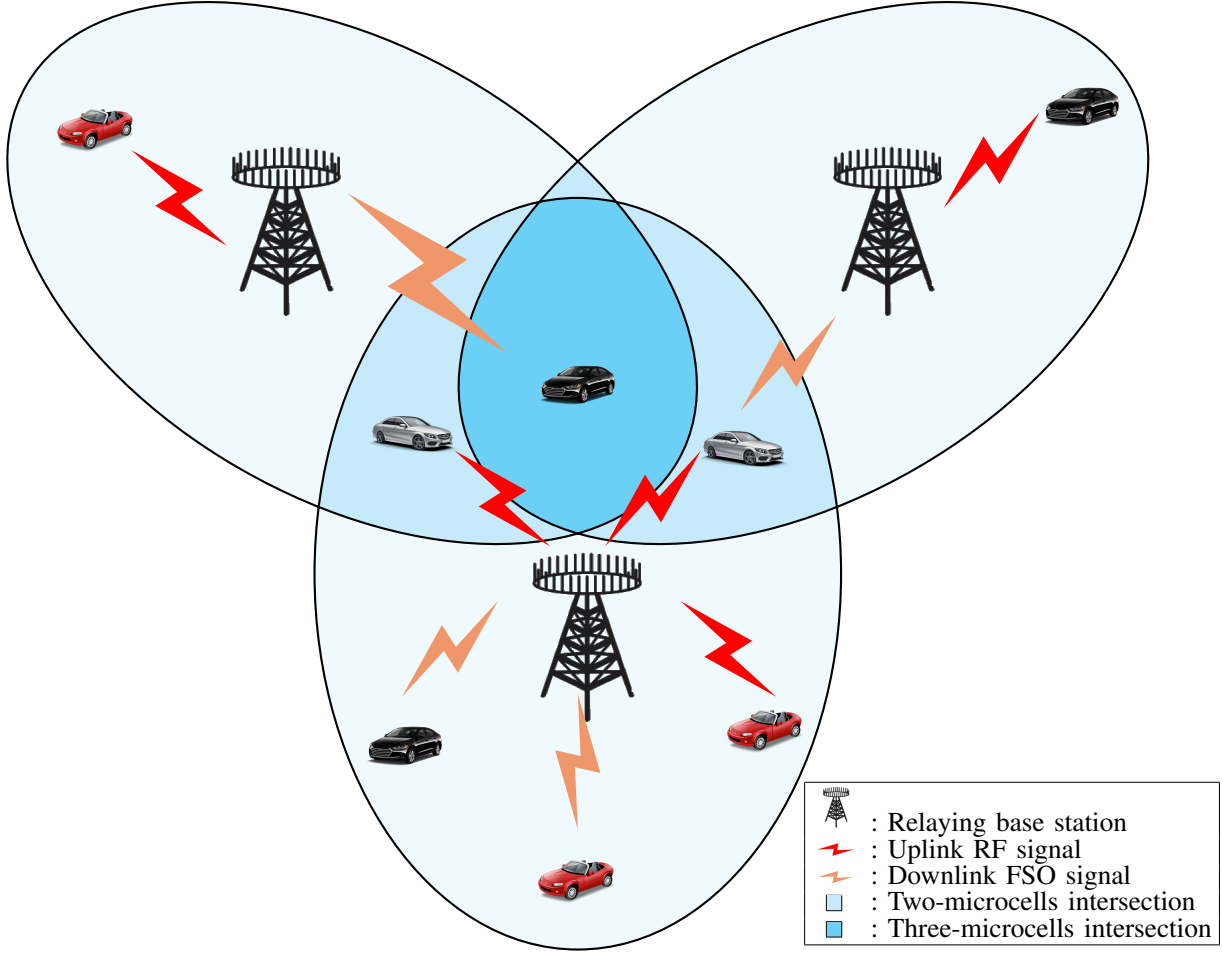}
    \caption{Scenario of outdoor vehicular communications of mixed RF/FSO cooperative relaying system. The vehicles are communicating through the relays that convert the incoming RF signal to FSO one. The system can be viewed as a hybrid cellular network where the source and the destination are the vehicles and the relays are the bases stations.}
\end{figure}
\subsubsection*{\textbf{High Power Amplifier Non-linearities Models}}
Since the distortion created by the HPA non-linearities is not linear and so the analysis will be somewhat complex, we refer to the Bussgang linearization theory to linearize the distortion. This theory states that the output of the non-linear HPA circuit is a function of the linear scale parameter $\Omega$ of the input signal and a non-linear distortion $\varsigma$ uncorrelated with the input signal and modeled as a complex Gaussian random variable $\varsigma \backsim \mathcal{CN} (0,~\sigma^2_{\varsigma})$. According to \cite{28,ofdm}, the parameters $\Omega$ and $\sigma^2_{\varsigma}$ for SEL are given by \cite[Eq.~(17)]{27}
\begin{equation}\label{eq:3}
\begin{split}
    \Omega = 1 - \exp\left(-\frac{A^2_{\text{sat}}}{\sigma^2_{\text{r}}}\right) + \frac{\sqrt{\pi} A_{\text{sat}}}{2\sigma^2_{\text{r}}}~\text{erfc}\left(\frac{A_{\text{sat}}}{\sigma_{\text{r}}}\right),
    \sigma^2_{\varsigma} = \sigma^2_{\text{r}}~\left[1 - \exp\left(-\frac{A^2_{\text{sat}}}{\sigma^2_{\text{r}}}\right) - \Omega^2 \right],
\end{split}    
\end{equation}
For TWTA, $\Omega$ and $\sigma^2_{\varsigma}$ are given by \cite[Eq.~(18)]{27}
\begin{equation}\label{eq:4}
\begin{split}
&\Omega = \frac{A^2_{\text{sat}}}{\sigma^2_{\text{r}}}\left[1 + \frac{A^2_{\text{sat}}}{\sigma^2_{\text{r}}}e^{\frac{A^2_{\text{sat}}}{\sigma^2_{\text{r}}}}  + \text{Ei}\left(-\frac{A^2_{\text{sat}}}{\sigma^2_{\text{r}}} \right)   \right],\\&~\sigma^2_{\varsigma} = -\frac{A^4_{sat}}{\sigma^2_{\text{r}}}\left[\left(1 + \frac{A^2_{\text{sat}}}{\sigma^2_{\text{r}}}\right)e^{\frac{A^2_{\text{sat}}}{\sigma^2_{\text{r}}}}\text{Ei}\left(-\frac{A^2_{\text{sat}}}{\sigma^2_{\text{r}}} \right) + 1 \right]-\sigma^2_{\text{r}}\Omega^2,
\end{split}
\end{equation} 
where $A_{\text{sat}}$, $\sigma^2_{\text{r}}$, $\text{erfc}(\cdot)$ and $\text{Ei}(\cdot)$ are the input saturation amplitude of the power amplifier, the mean power of the signal at the output of the gain block, the complementary error function, and the exponential integral function, respectively.\\
We also provide the expressions of the clipping factor for SEL and TWTA \cite[Eqs.~(13),~(14)]{28} as follows
\begin{equation}\label{eq:5}
\begin{split}
\eta_{\text{SEL}} = 1 - \exp\left(-\frac{A^2_{\text{sat}}}{\sigma^2_{\text{r}}}\right),~
\eta_{\text{TWTA}} = -\frac{A^4_{sat}}{\sigma^4_{\text{r}}}\left[\left(1+\frac{A^2_{\text{sat}}}{\sigma^2_{\text{r}}}\right)\exp\left(\frac{A^2_{\text{sat}}}{\sigma^2_{\text{r}}}\right)\text{Ei}\left(-\frac{A^2_{\text{sat}}}{\sigma^2_{\text{r}}} \right) + 1 \right],
\end{split}
\end{equation}
We also define the so-called the input back-off (IBO) given by $\text{IBO} = \frac{A^2_{\text{sat}}}{\sigma^2}$. Fig.~(2) illustrates the amplitude to amplitude (AM/AM) characteristics of the SEL, and TWTA with respect to the normalized input modulus for $A_{\text{sat}} = 1$.

\begin{figure}[H]
\centering
\setlength\fheight{5cm}
\setlength\fwidth{10cm}
\input{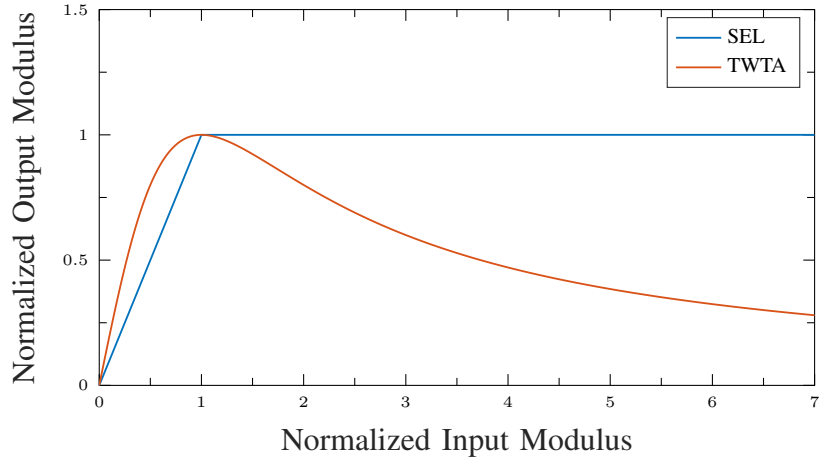}
    \caption{AM/AM characteristics of SEL, and TWTA}
\end{figure}
\subsection{Channels Models}
\subsubsection{\textbf{Statistics of RF channels}}
Since the RF channels experience Rayleigh fading, the PDF expression of the instantaneous SNR $\gamma_{1(m)}$ of the channel between $S$ and $R_{(m)}$ taking into account the outdated CSIs and the CSIs sorting can be written as follows \cite[Eq.~(8)]{66}
\begin{align}\label{eq:6}
\begin{split}
f_{\gamma_{1(m)}}(x) = m{N \choose m}\sum_{n=0}^{m-1} \frac{(-1)^n}{[(N-m+n)(1-\rho)+1]\overline{\gamma}_1}{m-1 \choose n}e^{-\frac{ (N-m+n+1)x }{[(N-m+n)(1-\rho)+1]\overline{\gamma}_1}},
\end{split}
\end{align}
After integrating the above expression, the CDF of $\gamma_{1(m)}$ can be obtained as
\begin{equation}\label{eq:7}
\begin{split}
F_{\gamma_{1(m)}}(x) = 1 - m{N \choose m}\sum_{n=0}^{m-1}{m-1 \choose n} \frac{(-1)^n}{N-m+n+1}e^{-\frac{ (N-m+n+1)x }{[(N-m+n)(1-\rho)+1]\overline{\gamma}_1}},
\end{split}
\end{equation}
Using the identity \cite[Eq.~(3.326.2)]{64}, the $t$-th moment of $\gamma_{1(m)}$ can be derived the  as follows
\begin{equation}\label{eq:8}
\begin{split}
\e{\gamma_{1(m)}^{t}} = m{N \choose m}\sum_{n=0}^{m-1} {m-1 \choose n}\Gamma(t+1) \frac{(-1)^n([(N-m+n)(1-\rho)+1]\overline{\gamma}_1)^{t}}{(N-m+n+1)^{t+1}},
\end{split}
\end{equation}
\subsubsection{\textbf{Statistics of FSO channels}}
The FSO part consists of three components $I_a, I_l$, and $I_p$ which are turbulence-induced fading, the path loss and the pointing error fading, respectively. The channel gain $I_m$ of the FSO between the relay $R_{(m)}$ and $D$ can be expressed as follows
\begin{equation}\label{eq:9}
I_m = I_a \cdot I_l \cdot I_p, 
\end{equation}
The table below summarizes the parameters of the optical part.
\vspace*{-1cm}
\begin{table}[H]
\renewcommand{\arraystretch}{1}
\caption{\textsc{FSO Sub-System}}
\label{tab:example}
\centering
\begin{tabular}{|c|c|}
    \hline
   \bfseries Parameter & \bfseries Definition \\
    \hline
     $\sigma$ &Weather attenuation\\
    \hline
    $\sigma^2_{\text{s}}$ & Jitter variance   \\
    \hline
    $\sigma^2_{\text{R}}$ & Rytov variance\\
    \hline
    $k$ & Wave number   \\
    \hline
    $\lambda$ & Wavelength  \\
    \hline
    $\xi$ & Pointing error coefficient   \\
    \hline
    $\omega_0$ & Beam waist at the relay\\
    \hline
    $\omega_{z}$ & Beam waist\\
    \hline
    $\omega_{z_{eq}}$ & Equivalent beam waist   \\
    \hline
    $L$ & Length of the optical link   \\
    \hline
    $a$ & Radius of the receiver aperture   \\
    \hline
    $A_0$ & Fraction of the collected power at $\text{L} = 0$  \\
    \hline
    $F_0$ & Radius of curvature\\
    \hline
    $C_n^2$ & Refractive index of the medium\\
    \hline   
    $R$ & Radial displacement of the beam at the receiver\\
    \hline
\end{tabular}
\end{table}

Using the Beers-Lambert law, the path loss can be expressed as follows \cite[Eq.~(12)]{49}
\begin{equation}\label{eq:10}
I_l = \exp(-\sigma L),
\end{equation}
The pointing error $I_p$ made by Jitter can be given as \cite[Eq.~(9)]{ln1}
\begin{equation}\label{eq:11}
I_p = A_0 \exp\left(-\frac{2R^2}{\omega^2_{z_{eq}}} \right),
\end{equation}
Assuming that the radial displacement of the beam at the detector follows the Rayleigh distribution, the PDF of the pointing error can be expressed as follows
\begin{equation}\label{eq:12}
f_{I_p}(I_p) = \frac{\xi^2}{A_0^{\xi^2}}I^{\xi^2-1}_p~,~~0\leq I_p\leq A_0,
\end{equation}
The pointing error coefficient can be expressed in terms of the Jitter standard deviation and the equivalent beam waist as follows
\begin{equation}\label{eq:13}
\xi = \frac{\omega_{z_{eq}}}{2\sigma_{\text{s}}},
\end{equation}
We can also relate $\omega_{z_{eq}}$ with the beam width $\omega_z$ of the Gaussian laser beam at the distance $L$ as follows
\begin{equation}\label{eq:14}
\omega^2_{z_{eq}} = \frac{\omega^2_{\text{L}}\sqrt{\pi}\text{erf}(v)}{2v\exp(-v^2)},
\end{equation}
where $v =\frac{\sqrt{\pi}a}{\sqrt{2}\omega_{\text{L}}}$, and \text{erf}($\cdot$) is the error function. The fraction of the collected power $A_0$ at the relay is given by
\begin{equation}\label{eq:15}
A_0 = |\text{erf}(v)|^2,
\end{equation}
The Gaussian beam waist can be defined as
\begin{equation}\label{eq:16}
\begin{split}
\omega_{z} = \omega_0\sqrt{(\Theta_0 + \Lambda_0)(1 + 1.63~\sigma_{\text{R}}^{12/5}\Lambda_1)},
\end{split}
\end{equation}
where $\Theta_0 = 1 - \frac{L}{F_0},~\Lambda_0 = \frac{2L}{kw_0^2},~\Lambda_1 = \frac{\Lambda_0}{\Theta_0^2 + \Lambda^2_0}$, and $\sigma_{\text{R}}^2$ is the Rytov variance given by \cite[Eq.~(15)]{49}
\begin{equation}\label{eq:17}
\sigma^2_{\text{R}} = 1.23~C^2_nk^{7/6}L^{11/6},
\end{equation}
The turbulence-induced fading $I_a$ is modeled by the Double Generalized Gamma and can be expressed as the product of two independent random variables $I_x$ and $I_y$ describing the large-scale and small-scale fluctuations, respectively. $I_x$ and $I_y$ each follows the generalized gamma distribution $I_x  \backsim GG(\alpha_1, m_1, \Omega_1)$ and $I_y  \backsim GG(\alpha_2, m_2, \Omega_2)$, where $m_1$ and $m_2$ are the shaping parametes defining the atmospheric turbulence fading. Moreover, $\alpha_1, \alpha_2, \Omega_1$ and $\Omega_2$ are defined using the variance of the small and large scale fluctutaions from \cite[Eqs.~(8.a), (8.b), and (9)]{dgg}. Thereby, the PDF of the turbulence-induced fading $I_a$ can be given by \cite[Eq.~(4)]{dgg}
\begin{align}\begin{split} 
{f_{I_a}}\left({I_a} \right) = \frac{\alpha _{2}{p^{{m_2} + \frac{1}{2}}}{q^{{m_1} - \frac{1}{2}}}{\left({2\pi } \right)}^{1 - \frac{p + q}{2}}}{\Gamma \left({m_1} \right)\Gamma \left({m_2} \right){I_a}}G_{p + q,0}^{0,p + q}\left(\frac{{p^p}{q^q}\Omega _1^q\Omega _2^p}{{m_1^q}{m_2^p}I_a^{\alpha_2p}} \left\vert\begin{array}{c} {\Delta \left({q{:}1 - {m_1}} \right)},{\Delta \left({p{:}1 - {m_2}} \right)}\\ {-}\end{array}\right. \!\!\right),\end{split}\end{align}
where $G^{m,n}_{p,q}[\cdot]$ is the Meijer's-G function, $p$ and $q$ are positive integers satisfying $\frac{p}{q} = \frac{\alpha_1}{\alpha_2}$ and $\Delta(j~;~x) \delequal \frac{x}{j}, \ldots, \frac{x + j - 1}{j}$. In case of the heterodyne detection, the average SNR $\mu_1$ is given by $\mu_1 = \frac{\eta\e{I_m}}{\sigma^2_0}$. Regarding the IM/DD detection, the average electrical SNR $\mu_2$ is given by $\mu_2 = \frac{(\eta\e{I_m})^2}{\sigma^2_0}$ while the instantaneous optical SNR is $\gamma_{2(m)} = \frac{(\eta I^2_m)}{\sigma^2_0}$. Unifying the two detection schemes and applying the transformation of the random variable $\gamma_{2(m)} = \frac{(\eta I_m)^r}{\sigma_0^2}$, the unified PDF of the instantaeous SNR $\gamma_{2(m)}$ can be expressed as follows
\begin{equation}\label{eq:19}
\begin{split}
f_{\gamma_{2(m)}}(\gamma) = \frac{\xi^2p^{m_2-\frac{1}{2}}q^{m_1-\frac{1}{2}}(2\pi)^{1-\frac{p+q}{2}}}{r\Gamma(m_1)\Gamma(m_2)\gamma} G_{p+q+\alpha_2p,\alpha_2p}^{0,p+q+\alpha_2p} \Bigg(\left(\frac{p\Omega_1}{m_2}\right)^p \left(\frac{q\Omega_2}{m_1}\right)^q \left(\frac{\mu_r(A_0 I_l)^{r}}{\gamma}\right)^{\frac{\alpha_2p}{r}} ~\bigg|~\begin{matrix} \kappa_1 \\ \kappa_2 \end{matrix} \Bigg),
\end{split}
\end{equation}
where $\sigma_0^2$, $\eta$ are the channel noise and the electrical-to-optical conversion coefficient, respectively. The parameter $r$ takes two values 1 and 2 standing for heterodyne and IM/DD, respectively. The vectors $\kappa_1$ and $\kappa_2$ are given by
\begin{equation}
\begin{split}
\kappa_1 = \Delta(\alpha_2p:1-\xi^2),~\Delta(q:1 - m_1),~\Delta(p:1 - m_2),~\kappa_2 = \Delta(\alpha_2p:-\xi^2),
\end{split}
\end{equation}
The average SNR $\overline{\gamma}_r$\footnote[1]{The average SNR $\overline{\gamma}_r$ is defined as $\overline{\gamma}_r = \eta^r\e{I_{m}^r}/\sigma_{0}^2$, while the average electrical SNR $\mu_r$ is given by $\mu_r = \eta^r\e{I_{m}}^r/\sigma_{0}^2$. Therefore, the relation between the average SNR and the average electrical SNR is trivial given that $ \frac{\e{I^2_{m}}}{\e{I_{m}}^2} = \sigma^2_{\text{si}} + 1$, where $\sigma^2_{\text{si}}$ is the scintillation index \cite{scin}.} can be expressed as
\begin{align}
 \overline{\gamma}_r = \frac{\e{I^r_{m}}}{\e{I_{m}}^r}\mu_r,   
\end{align}
where $\mu_r$ is the average electrical SNR given by
\begin{equation}
    \mu_r = \frac{\eta^r\e{I_{m}}^r}{\sigma_{0}^2},
\end{equation}
After integrating Eq.~(19), the CDF of the instantaneous SNR $\gamma_{2(m)}$ can be expressed as follows
\begin{equation}\label{eq:21}
\begin{split}
F_{\gamma_{2(m)}}(\gamma) = \frac{\xi^2p^{m_2-\frac{3}{2}}q^{m_1-\frac{1}{2}}(2\pi)^{1-\frac{p+q}{2}}}{\alpha_2\Gamma(m_1)\Gamma(m_2)}  G_{p+q+2\alpha_2p,2\alpha_2p}^{\alpha_2p,p+q+\alpha_2p} \Bigg(\left(\frac{p\Omega_1}{m_2}\right)^p \left(\frac{q\Omega_2}{m_1}\right)^q \left(\frac{\mu_r(A_0 I_l)^{r}}{\gamma}\right)^{\frac{\alpha_2p}{r}}\bigg|\begin{matrix} \kappa_3 \\ \kappa_4 \end{matrix} \Bigg),
\end{split}
\end{equation}
The vectors $\kappa_3$ and $\kappa_4$ are given by
\begin{equation}
\begin{split}
\kappa_3 = \Delta(\alpha_2p:1-\xi^2),~\Delta(q:1 - m_1),~\Delta(p:1 - m_2),~[1]_{\alpha_2p},~\kappa_4 = [0]_{\alpha_2p},~\Delta(\alpha_2p:-\xi^2),
\end{split}
\end{equation}
where $[x]_j$ is defined as the vector of length $j$ and its components are equal to $x$.\\
After changing the variable of the integration ($x = \gamma^{-\frac{\alpha_2p}{r}
}$) and applying the following identity \cite[Eq.~(2.24.2.1)]{62}, the $t$-th moment of the optical SNR can be derived as follows
\begin{equation}\label{eq:22}
\begin{split}
\e{\gamma_{2(m)}^t} = \ddfrac{\xi^2p^{m_2-1}q^{m_1-\frac{1}{2}}(2\pi)^{1-\frac{p+q}{2}}\zeta^{t\left[\frac{r}{\alpha_2p}-1\right]-1}}{\Gamma(m_1)\Gamma(m_2)\prod_{j=1}^{\alpha_2p}\Gamma\left(t\left[\frac{r}{\alpha_2p}-1\right]-\kappa_{2,j}\right)}
\frac{\prod_{j=1}^{p+q+\alpha_2p}\Gamma\left(t\left[\frac{r}{\alpha_2p}-1\right]-\kappa_{1,j}\right)}{\prod_{j=p+q+\alpha_2p+1}^{p+q+2\alpha_2p}\Gamma\left(t\left[\frac{r}{\alpha_2p}-1\right]-\kappa_{1,j}\right)},
\end{split}
\end{equation}
where $\zeta = \left(\frac{p\Omega_1}{m_2}\right)^p \left(\frac{q\Omega_2}{m_1}\right)^q (A_0 I_l)^{\alpha_2p}\mu_r^{\frac{\alpha_2p}{r}}$.
\section{Performance Analysis of Fixed Gain Relaying}
This relaying scheme consists of amplifying the signal by a fixed gain based on the average received CSI. The gain factor can be expressed as follows
\begin{equation}\label{eq:23}
G = \sqrt{\frac{\sigma^2_{\text{r}}}{\e{|h_{1(m)}|^2}P_1+\sigma^2_0}},
\end{equation}
where $P_1$ is the average transmitted power from $S$. The end-to-end Signal-to-Noise-plus-Distortion-Ratio (SNDR) can be expressed as follows \cite[Eq.~(16)]{28}
\begin{equation}\label{eq:24}
\gamma_{\text{e2e}} = \frac{\gamma_{1(m)} \gamma_{2(m)}}{\kappa\gamma_{2(m)} + \e{\gamma_{1(m)}} + \kappa}, 
\end{equation}
The HPA non-linearities factor $\kappa$ can be given by \cite[Eq.~(17)]{28}
\begin{equation}\label{eq:25}
\kappa = 1 + \frac{\sigma^2_{\varsigma}}{\Omega^2G^2\sigma^2_0},
\end{equation}
Note that for the case of linear relaying, the factor $\kappa$ is reduced to one and so the end-to-end SNR (27) describes an ideal system.
\subsection{Outage Probability Analysis}
The outage probability (OP) is defined as the probability that the end-to-end SNDR falls below a given threshold $\gamma_{\text{th}}$. It can be generally written as
\begin{equation}\label{eq:26}
P_{\text{out}}(\gamma_{\text{th}}) \delequal \text{Pr}[\gamma_{\text{e2e}} < \gamma_{\text{th}}] = F_{\gamma_{\text{e2e}}}(\gamma_{\text{th}}),
\end{equation} 
where $F_{\gamma_{\text{e2e}}}(\cdot)$ is the CDF of the end-to-end SNDR. After substituting (27) in (29), the OP can be derived as follows
\begin{equation}\label{eq:27}
\begin{split}
P_{\text{out}}(\gamma_{\text{th}}) =& 1 - \frac{m\xi^2p^{m_2-1}q^{m_1-\frac{1}{2}}r^{\mu - 1}}{\sqrt{\alpha_2}~\Gamma(m_1)\Gamma(m_2)(2\pi)^{\frac{\alpha_2p + r(p+q) - 3}{2}}} {N \choose m} \sum_{n = 0}^{m-1}{m-1 \choose n}\frac{(-1)^n}{N-m+n+1}\\& \times~ \exp(-\beta\kappa\gamma_{\text{th}})
 G_{r(p+q+\alpha_2p)+\alpha_2p,r\alpha_2p}^{0,r(p+q+\alpha_2p)+\alpha_2p} \Bigg[\left(\frac{\alpha_2p}{\beta\gamma_{\text{th}}c}\right)^{\alpha_2p}(\zeta r^{p+q})^r~\bigg|~\begin{matrix} \kappa_5 \\ \kappa_6 \end{matrix} \Bigg],
 \end{split}
\end{equation}
where $\beta = \frac{N-m+n+1}{[(N-m+n)(1-\rho)+1]\overline{\gamma}_1}$, $\mu = \sum_{j=1}^{\alpha_2p}\kappa_{2,j} - \sum_{j=1}^{p+q+\alpha_2p}\kappa_{1,j} + \frac{p+q}{2} + 1, c= (\kappa + 1)\e{\gamma_{1(m)}}$, and the vectors $\kappa_5, \kappa_6$ are given by
\begin{equation}
\begin{split}
\kappa_5 = [1]_{\alpha_2p},~\Delta(r:\alpha_2p:1-\xi^2),~\Delta(r:q:1-m_1),~\Delta(r:p:1-m_2),~\kappa_6 = \Delta(r:\alpha_2p:-\xi^2),
\end{split}
\end{equation}
The operator $\Delta(\cdot:\cdot:\cdot)$ is defined by $\Delta(r:j:x) \delequal \Delta\left(r:\frac{x}{j}\right),\ldots,~\Delta\left(r:\frac{x+j-1}{j}\right)$.
\begin{proof}
The proof of Eq.~(30) is given in Appendix A.
\end{proof}
We also derive the asymptotic high SNR using the expansion of the Meijer's-G function for large values of the average electrical SNR $\mu_r$ as follows
\begin{equation}\label{eq:28}
\begin{split}
&P^{\infty}_{\text{out}}(\gamma_{\text{th}}) \underset{\mu_r \gg 1}\cong 1 - \frac{m\xi^2p^{m_2-1}q^{m_1-\frac{1}{2}}r^{\mu - 1}}{\sqrt{\alpha_2}~\Gamma(m_1)\Gamma(m_2)(2\pi)^{\frac{\alpha_2p + r(p+q) - 3}{2}}} {N \choose m} \sum_{n = 0}^{m-1}{m-1 \choose n}\frac{(-1)^n}{N-m+n+1}\\& \times~\exp(-\beta\kappa\gamma_{\text{th}}) ~\sum_{i=1}^{r(p+q+\alpha_2p)+\alpha_2p}
\frac{\prod_{j=1,~j \neq i}^{r(p+q+\alpha_2p)+\alpha_2p}\Gamma(\kappa_{5,i}-\kappa_{5,j})}{\prod_{j=1}^{r\alpha_2p}\Gamma(\kappa_{5,i}-\kappa_{6,j})}\left[\left(\frac{\alpha_2p}{\beta \gamma_{\text{th}}c}\right)^{\alpha_2p} (\zeta r^{p+q})^r   \right]^{\kappa_{5,i}-1},
 \end{split}
\end{equation}
Eq.~(32) provides engineering insights about the achieved gain such as the diversity order $G_d$. Note that the system saturated at high SNR since the impact of the hardware impairments becomes more pronounced at high rate, and hence an outage floor is created. Consequently the system achieves no gain $G_d = 0$. In the absence of the hardware impairments, the system achieves a diversity gain equal to
\begin{equation}
    G_d = \min\left(1,~\frac{\alpha_1m_1}{r},~\frac{\alpha_2m_2}{r},\frac{\xi^2}{r}\right),
\end{equation}
Note that in our previous work \cite{27}, the system employs the opportunistic relay selection protocol with outdated CSI. We proved that the diversity gain achieved is equal to $G_d = N$ for full correlation ($\rho = 1$) and $G_d = 1$ for outdated CSI ($\rho < 1$). Since this proposed system employs partial relay selection, however, the diversity gain for the RF sub-system is always $G_d = 1$ for either perfect or outdated CSI. 
\vspace*{-1cm}
\subsection{Bit Error Probability Analysis}
The bit error probability (BEP) expression can be given by
\begin{equation}
\overline{P_e} = \frac{\delta}{2\Gamma(\tau)}\sum_{k=1}^{v}\int\limits_{0}^{\infty}\Gamma(\tau,q_k\gamma)f_{\gamma_{\text{e2e}}}(\gamma)d\gamma,
\end{equation}
where $v,~\delta,~\tau$~, and $q_k$ vary depending on the type of
detection (heterodyne technique or IM/DD) and modulation
being assumed. It is worth accentuating that this expression is
general enough to be used for both heterodyne and IM/DD
techniques and can be applicable to different modulation
schemes. The parameters $v,~\delta,~\tau$~, and $q_k$ are summarized in Table II.
\vspace*{-0.5cm}
\begin{table}[H]
\renewcommand{\arraystretch}{1.3}
\caption{\textsc{Parameters for Different Modulations$^\dag$}}
\label{tab:example}
\centering
\begin{tabular}{|c|c|c|c|c|c|}
    \hline
    \textbf{Modulation}  &  $\delta$ & $\tau$ & $q_k$ & $v$ & \textbf{Detection}\\
    \hline
   \textbf{OOK}    &  1& 0.5 & 0.5 & 1 & IM/DD\\
    \hline
   \textbf{BPSK}    &  1& 0.5 & 1 & 1 & Heterodyne\\
    \hline
     \textbf{M-PSK}    &  $\frac{2}{\max(\log_2(M), 2)}$ & 0.5 & $\text{sin}^2\left(\frac{(2k-1)\pi}{M}\right)$ & $\max(\frac{M}{4}),1$ & Heterodyne\\
    \hline
      \textbf{M-QAM}    & $\frac{4}{\log_2(M)}\left(1-\frac{1}{\sqrt{M}}  \right)$  & 0.5 & $\frac{3(2k-1)^2}{2(M-1)}$& $\frac{\sqrt{M}}{2}$& Heterodyne \\
    \hline
\end{tabular}
\\ 
\rule{0in}{1.2em}$^\dag$\scriptsize In case of OOK modulation, the parameters $v,~\delta,~\tau$~, and $q_k$ are given by \cite[Eq.~(26)]{mod1}. For M-PSK and M-QAM modulations, these parameters are provided by \cite[Eqs.(30), (31)]{50}.
\end{table}
\vspace*{-1cm}
The average bit error probability expression in (34) can be rewritten in terms of the CDF by using integration by parts as
\begin{equation}
\overline{P_e} = \frac{\delta}{2\Gamma(\tau)}\sum_{k=1}^{v}q_k^{\tau}\int\limits_{0}^{\infty}\gamma^{\tau-1}e^{-q_k\gamma}F_{\gamma_{\text{e2e}}}(\gamma)d\gamma,    
\end{equation}
First, we should replace the expression of the CDF of the end-to-end SNDR (30) in (35) and then we invert the argument of the Meijer's-G function. After transforming the exponential into Meijer's-G function \cite[Eq.~(07.34.03.0046.01)]{68} and using the identity \cite[Eq.~(2.24.1)]{62}, the BEP is finally derived as follows
\begin{equation}
\begin{split}
&\overline{P_e} = \frac{\delta v}{2} - \ddfrac{m\delta\xi^2p^{\tau+m_2-\frac{3}{2}}q^{m_1-\frac{1}{2}}\alpha_2^{\tau-\frac{3}{2}}r^{\mu-1}}{2(2\pi)^{\frac{2\alpha_2p+r(p+q)-4}{2}}\Gamma(m_1)\Gamma(m_2)\Gamma(\tau)} {N \choose m} \sum_{n=0}^{m-1}\sum_{k=1}^v {m-1 \choose n} \frac{(-1)^n}{N-m+n+1}\\&\times\left(\frac{q_k}{\beta\kappa+q_k}\right)^{\tau}
G_{(r+1)\alpha_2p,r(p+q+\alpha_2p)+\alpha_2p}^{r(p+q+\alpha_2p)+\alpha_2p,\alpha_2p} \Bigg(\left(\frac{\beta c}{\beta\kappa + q_k}\right)^{\alpha_2p}(\zeta r^{p+q})^{-r}~\bigg|~\begin{matrix} \Delta(\alpha_2p,1-\tau),~1-\kappa_6 \\ 1-\kappa_5 \end{matrix} \Bigg),
\end{split}    
\end{equation}
\vspace*{-1cm}
\subsection{Ergodic Capacity Analysis}
The system capacity, expressed in bps/Hz, is defined as the maximum error-free data rate transferred by the system channel. It can be expressed as follows
\begin{equation}\label{eq:32}
\overline{C} \delequal \e{\log_{2}(1+\varpi\gamma)},
\end{equation}
where $\varpi$ can take the values 1 or $e/2\pi$ for heterodyne or IM/DD, respectively. 
After some mathematical manipulations, the ergodic capacity can be expressed in terms of the complementary CDF $\overline{F}_{\gamma}$ as follows
\begin{equation}\label{eq:33}
\overline{C} = \frac{\varpi}{\text{ln(2)}}\int\limits_{0}^{\infty} (1+\varpi \gamma)^{-1} \overline{F}_{\gamma_{\text{e2e}}}(\gamma)~d\gamma,
\end{equation}
After replacing the complementary CDF of (30) in Eq.~(38), then we should transform $(1+\varpi \gamma)^{-1}$, the exponential and the Meijer's-G into the Fox-H function. Applying the following identity \cite[Eq.~(2.3)]{63} and after some mathematical manipulations, the average ergodic capacity can be derived as follows
\begin{equation}\label{eq:34}
\begin{split}
&\overline{C} = \frac{m\xi^2p^{m_2-\frac{1}{2}}q^{m_1-\frac{1}{2}}\varpi r^{\mu - 1}}{(\alpha_2p)^{\frac{3}{2}}(2\pi)^{\frac{\alpha_2p+r(p+q)-3}{2}}\text{ln(2)}\Gamma(m_1)\Gamma(m_2)\kappa}{N \choose m}\sum_{n=0}^{m-1}{m-1 \choose n}\frac{(-1)^n}{(N-m+n+1)\beta} \\&H_{1,0:1,1:r(p+q+\alpha_2p)+\alpha_2p,r\alpha_2p}^{0,1:1,1:0,r(p+q+\alpha_2p)+\alpha_2p}\left(\begin{matrix}(0;1,-1)\\-\end{matrix}\bigg|\begin{matrix}(0,1)\\(0,1)\end{matrix}\bigg|\begin{matrix}(\kappa_5,[\frac{1}{\alpha_2p}]_{r(p+q+\alpha_2p)+\alpha_2p})\\(\kappa_6,[\frac{1}{\alpha_2p}]_{r\alpha_2p})\end{matrix}\bigg|\frac{\varpi}{\beta \kappa},\frac{\alpha_2p \kappa}{c}(\zeta r^{p+q})^{\frac{r}{\alpha_2p}}\right),
\end{split}
\end{equation}
where $H_{p_1,q_1:p_2,q_2:p_3,q_3}^{m_1,n_1:m_2,n_2:m_3,n_3}[\cdot,\cdot]$ is the bivariate Fox-H function. An efficient MATLAB implementation of this function is given in \cite[Appendix(B)]{bfox}.\\ 
Since the relays are impaired, we can also derive a ceiling in terms of the impairment clipping factor that limits the capacity as the impairment becomes more severe. This ceiling is given by \cite[Eq.~(37)]{28}
\begin{equation}\label{eq:35}
\overline{C_c} = \log_2\left(1 + \frac{\varpi \Omega^2}{\eta_{\text{SEL/TWTA}}-\Omega^2}  \right),
\end{equation} 
\section{performance Analysis of Variable Gain Relaying}
This relaying scheme consists of amplifying the signal by a variable gain based on the instantaneous received CSI. The gain factor can be written as follows
\begin{equation}\label{eq:36}
G = \sqrt{\frac{\sigma^2_{\text{r}}}{|h_{1(m)}|^2P_1+\sigma^2_0}},
\end{equation}
The end-to-end SNDR can be formulated as follows \cite[Eq.~(14)]{32}
\begin{equation}\label{eq:37}
\gamma_{\text{e2e}} = \frac{\gamma_{1(m)}\gamma_{2(m)}}{\kappa\gamma_{2(m)} + \gamma_{1(m)} + \kappa},
\end{equation}
The closed-form of the end-to-end SNDR statistics in (42) is mathematically intractable. Thereby, we consider an approximate expression of the end-to-end SNDR as follows
\begin{equation}\label{eq:38}
\gamma_{\text{e2e}} \cong \text{min}\left(\gamma_{1(m)},~\frac{\gamma_{2(m)}}{(\kappa-1) \gamma_{2(m)} + 1}\right),
\end{equation}
\subsection{Outage Probability Analysis}
Since the derivation of the OP is intractable, we derive a tight upper bound based on (43) as follows
\begin{equation}\label{eq:39}
\begin{split}
P^{\text{up}}_{\text{out}}(\gamma_{\text{th}}) = F_{\gamma_{1(m)}}(\gamma_{\text{th}}) + F_{\gamma_{2(m)}}\left(\frac{\gamma_{\text{th}}}{(\kappa-1) \gamma_{\text{th}} + 1}\right) - F_{\gamma_{1(m)}}(\gamma_{\text{th}})F_{\gamma_{2(m)}}\left(\frac{\gamma_{\text{th}}}{(\kappa-1) \gamma_{\text{th}} + 1}\right),
\end{split}
\end{equation}
To get a deep scope about the system behavior, we derive an asymptotic high SNR using the Meijer's-G expansion of the CDF of $\gamma_{2(m)}$ as follows
\begin{equation}\label{eq:40}
\begin{split}
&G_{p+q+2\alpha_2p,2\alpha_2p}^{\alpha_2p,p+q+\alpha_2p} \Bigg(\zeta \left( \frac{1 + (\kappa-1)\gamma_{\text{th}}}{\gamma_{\text{th}}} \right)^{\frac{\alpha_2p}{r}} ~\bigg|~\begin{matrix} \kappa_3 \\ \kappa_4 \end{matrix} \Bigg) \underset{\mu_r \gg 1} \cong   \sum_{i=1}^{p+q+\alpha_2p}\left[ \zeta \left( \frac{1 + (\kappa-1)\gamma_{\text{th}}}{\gamma_{\text{th}}} \right)^{\frac{\alpha_2p}{r}}\right]^{\kappa_{3,i}-1} \\&\times~\frac{\prod_{j=1,~j \neq i}^{p+q+\alpha_2p}\Gamma(\kappa_{3,i} - \kappa_{3,j})\prod_{j=1}^{\alpha_2p}\Gamma(1-\kappa_{3,i}+\kappa_{4,j})}{\prod_{j=\alpha_2p+1}^{2\alpha_2p}\Gamma(\kappa_{3,i}-\kappa_{4,j})\prod_{j=p+q+\alpha_2p+1}^{p+q+2\alpha_2p}\Gamma(\kappa_{3,j}-\kappa_{3,i}+1)},
\end{split}
\end{equation}
\vspace*{-1cm}
\subsection{Bit Error Probability Analysis}
Since a closed-form of the BEP derived by introducing the upper bound (44) in (35) is not solvable due to the impairment factor, we can only derive an asymptotic high SNR expression of the BEP. At high SNR, the CDF of the overall SNDR can be approximated as follows
\begin{equation}
F_{\gamma_{\text{e2e}}}(\gamma) \underset{\mu_r \gg 1} \cong F_{\gamma_{1(m)}}(\gamma) + F_{\gamma_{2(m)}}\left(\frac{\gamma}{(\kappa-1) \gamma + 1}\right),
\end{equation}
\begin{equation}
\begin{split}
&\overline{P}_e  \underset{\mu_r \gg 1} \cong \frac{v\delta}{2} - \frac{m\delta}{2\Gamma(\tau)}{N \choose m} \sum_{k=1}^v\sum_{n=0}^m-1 {m-1 \choose n}\frac{(-1)^n}{N-m+n+1}\left(\frac{q_k}{\beta + q_k}\right)^{\tau}+\frac{\delta\xi^2}{2\alpha_2}\\&\times~\frac{p^{m_2-\frac{3}{2}}q^{m_1-\frac{1}{2}}(2\pi)^{1-\frac{p+q}{2}}}{\Gamma(m_1)\Gamma(m_2)\Gamma(\tau)}\sum_{k=1}^v\sum_{i=1}^{p+q+\alpha_2p}\frac{\prod_{j=1,~j \neq i}^{p+q+\alpha_2p}\Gamma(\kappa_{3,i} - \kappa_{3,j})\prod_{j=1}^{\alpha_2p}\Gamma(1-\kappa_{3,i}+\kappa_{4,j})}{\prod_{j=\alpha_2p+1}^{2\alpha_2p}\Gamma(\kappa_{3,i}-\kappa_{4,j})\prod_{j=p+q+\alpha_2p+1}^{p+q+2\alpha_2p}\Gamma(\kappa_{3,j}-\kappa_{3,i}+1)} \\&\times~\frac{q_k^{\frac{\alpha_2p}{r}(\kappa_{3,i}-1)}\zeta^{\kappa_{3,i}-1}}{\Gamma\left(\frac{\alpha_2p}{r}(\kappa_{3,i}-1)\right)}G_{2,1}^{1,2}  \left( \frac{\kappa-1}{q_k} ~\bigg|~\begin{matrix} 1- \frac{\alpha_2p}{r}(\kappa_{3,i}-1)-q_k,~1+\frac{\alpha_2p}{r}(\kappa_{3,i}-1) \\ 0 \end{matrix} \right),
\end{split}    
\end{equation}
After using the following identities \cite[Eq.~(3.351.3)]{64}, and \cite[Eq.~(2.24.3.1)]{62}, the BEP can be given by (47).
\vspace*{-1cm}
\subsection{System Gains}
For the most coherent linear modulation, the BEP can be reformulated as follows
\begin{equation}
\overline{P_e} = \e{\mathcal{Q}(\sqrt{c\gamma})},
\end{equation}
where $\mathcal{Q}(\cdot)$ is the Gaussian-$Q$ function, and $c$ is a parameter related to the format of the modulation, e.g, $c = 2$ stands for BPSK modulation. After applying an integration by parts on Eq.~(48), BEP can be written as
\begin{equation}
\overline{P_e} = \sqrt{\frac{c}{8\pi}}\int\limits_0^\infty\frac{e^{-\frac{c}{2}\gamma}}{\sqrt{\gamma}}F_{\gamma}(\gamma)d\gamma,    
\end{equation}
The derivation of the closed-form of the BEP is mathematically not tractable due to the presence of the terms related to the hardware impairments. Thereby, a numerical integration is needed. As we mentioned earlier, the hardware impairments introduces indesirable effects on the reliability of the system and this effects become more significant for high SNR range. As a result, an irreducuble floor is created and degrades the error performance as the transmitted power increases. Therefore, the diversity gain $G_d$ is equal to zero. Now, considering an ideal system and since the CDF of the instantaneous SNR consists of complex functions such as the Meijer-G function, such function did not unpack engineering insights about the system gains. Consequently, it is more meaningful to derive the BEP at high SNR range as follows
\begin{equation}
\overline{P_e} \approx (G_c \overline{\gamma})^{-G_d}, 
\end{equation}
where $G_d$ and $G_c$ are the diversity and the coding gains, respectively. To get this form of the BEP, we refer to the technique proposed by \cite{e1,e4} to approximate the PDF of the overall SNR as follows
\begin{equation}
f_{\gamma}(\gamma) = a\gamma^{b} + o(\gamma),    
\end{equation}
From the above approximation, the asymptotical high SNR expression of the BEP can be written as
\begin{equation}
\begin{split}
 \overline{P_e} \approx  \frac{\prod_{i=1}^{b+1}(2i-1)}{2(b+1)!c^{b+1}}\frac{\partial^b f_{\gamma}}{\partial \gamma^b}(0) = \frac{2^ba\Gamma(b+3/2)}{\sqrt{\pi}(b+1)}(c\overline{\gamma})^{-(b+1)},  
\end{split}
\end{equation}
where $a$ is a constant and $b$ must be a natural number for the first equation in (52) and not necessarily an integer for the second equation. Consequently, we derive the approximate expression of the PDF to find the diversity gain $G_d = b+1$ and the coding gain $G_c$. Given that the CDF of the overall SNR for the ideal case, it can be approximated at high SNR region as
\begin{equation}
    F_{\gamma}(\gamma) \approx F_{\gamma_{1(m)}}(\gamma) + F_{\gamma_{2(m)}}(\gamma),
\end{equation}
Deriving (53) gives the approximate PDF of the end-to-end SNR as
\begin{equation}
  f_{\gamma}(\gamma) \approx f_{\gamma_{1(m)}}(\gamma) + f_{\gamma_{2(m)}}(\gamma),  
\end{equation}
Since $\gamma_{1(m)}$ is exponentially distributed under the assumption of PRS with outdated CSI, $b$ is equal to zero. On the other side, the high SNR approximation of $f_{\gamma_{2(m)}}$ can be derived by using the expansion of the Meijer-G function given by Eq.~(45). Note that for Eq.~(46), we must substitute $\kappa = 1$ to consider the ideal case.\\
Consequently, the PDF of $\gamma_{2(m)}$ can be written as
\begin{equation}
f_{\gamma_{2(m)}}(\gamma)\approx D \gamma^{\min\left(\frac{\xi^2}{r},~\frac{\alpha_1m_1}{r},~\frac{\alpha_2m_2}{r}\right)},
\end{equation}
where D is a constant parameter. After combining the PDF approximations of $\gamma_{1(m)}$ and $\gamma_{2(m)}$, the PDF of the overall SNR can be derived as follows
\begin{equation}
    f_{\gamma}(\gamma) \approx a \gamma^{\min\left(1,~\min\left(\frac{\xi^2}{r},~\frac{\alpha_1m_1}{r},~\frac{\alpha_2m_2}{r}\right)\right)},
\end{equation}
Finally, the diversity gain $G_d$ can be given by
\begin{equation}
    G_d = \min\left(1,~\min\left(\frac{\xi^2}{r},~\frac{\alpha_1m_1}{r},~\frac{\alpha_2m_2}{r}\right)\right),
\end{equation}
While the coding gain $G_c$ can be derived as follows
\begin{equation}
G_c = c \left( \frac{2^ba\Gamma(a+3/2)}{\sqrt{\pi}(b+1)}  \right)^{-\frac{1}{b+1}},   
\end{equation}
\vspace*{-1cm}
\subsection{Ergodic Capacity Analysis}
The closed-form can be computed by numerical integration using the PDF of the end-to-end SNDR. However, deriving a closed-form of the channel capacity in our case is very complex if not impossible. To overcome this problem, we should refer to the approximation given by \cite[Eq.~(35)]{32}
\begin{equation}\label{eq:44}
\e{\log_{2}\left(1+\frac{\varphi}{\psi}\right)} \cong \log_{2}\left(1 + \frac{\e{\varphi}}{\e{\psi}}\right),
\end{equation}
Given that the RF and FSO channels are independent and using (44), we can derive an approximate expression of the ergodic capacity.\\
To characterize the ergodic capacity of our system, we derive an upper bound using the following theorem:
\newtheorem{theorem}{Theorem}
\begin{theorem}
For asymmetric (Rayleigh/Double Generalized Gamma) fading channels, the ergodic capacity $\overline{C}$ in (bps/Hz) with AF and non-linear relaying has an upper bound using the Jensen's inequality as follows
\begin{equation}\label{eq:45}
    \overline{C} \leq \log_{2}\left(1 + \varpi \mathcal{J} \right),
\end{equation}
\end{theorem}
The term $\mathcal{J}$ is given by Eq.~(61). The capacity ceiling $\overline{C}_c$ is the same as the FG relaying scheme (40).
\vspace*{-0.5cm}
\begin{proof}
The proof of Eq.~(61) is provided in Appendix B.
\end{proof}
\vspace*{-1cm}
\begin{equation}\label{eq:46}
\begin{split}
&\mathcal{J} = \frac{m\xi^2p^{m_2-1}q^{m_1-\frac{1}{2}}(2\pi)^{1-\frac{p+q}{2}}}{\alpha_2\kappa\Gamma(m_1)\Gamma(m_2)}{N \choose m}\sum_{n=0}^{m-1}{m-1 \choose n}\frac{(-1)^{n+1}}{[(N-m+n)(1-\rho)+1]\overline{\gamma}_1\beta^2}\\& \times
H_{1,0:0,2:p+q+\alpha_2p,\alpha_2p}^{0,1:2,0:0,p+q+\alpha_2p}\left(\begin{matrix} (-1;1,1) \\ - \end{matrix} ~\bigg|~  \begin{matrix} (1,1) \\ (-1,1),~(0,1) \end{matrix} ~\bigg|~ \begin{matrix} (\kappa_1,[-\frac{r}{\alpha_2p}]_{p+q+\alpha_2p}) \\ (\kappa_2,[-\frac{r}{\alpha_2p}]_{\alpha_2p})\end{matrix}~\bigg|~ -1,~ -\frac{\zeta^{-\frac{r}{\alpha_2p}}}{\beta \kappa}  \right),
\end{split}
\end{equation}
\vspace*{-1cm}
\section{Numerical Results and Discussion}
In this section, we verify the analylical expressions with the numerical results using the Monte Carlo simulation \footnote{For all cases, $10^9$ realizations of the random variables were generated to perform the Monte Carlo simulation in MATLAB.}. Temporally correlated Rayleigh channel coefficients are generated using (1). The atmospheric turbulence $I_a$ is generated using the expression $I_a = I_{aX}\times I_{aY}$, where the two independent random variables $I_{aX}$ and $I_{aY}$ follow the Generalized Gamma distribution using \cite{65}. In addition, the pointing error is simulated by generating the radial displacement $R$ following the Rayleigh distribution with scale equal to the jitter standard deviation ($\sigma_s$) and then we generate the samples using (11). Since the path loss is deterministic, it can be generated using the relation (10). Table III summarizes the main simulation parameters.
\vspace*{-1cm}
\begin{table}[H]
\renewcommand{\arraystretch}{1}
\caption{\textsc{Main Simulation Parameters}}
\label{tab:example}
\centering
\begin{tabular}{|c|c|}
    \hline
    Parameter  &  Value\\
    \hline
    $L$    &    1 km\\
    \hline
    $\lambda$    & 1550 nm   \\
    \hline
    $F_0$ & -10 m\\
    \hline
    $a$    & 5 cm  \\
    \hline
    $\omega_0$ & 5 mm\\
    \hline
    $\sigma_s$ & 3.75 cm\\
    \hline
\end{tabular}
\end{table}
\vspace*{-0.5cm}
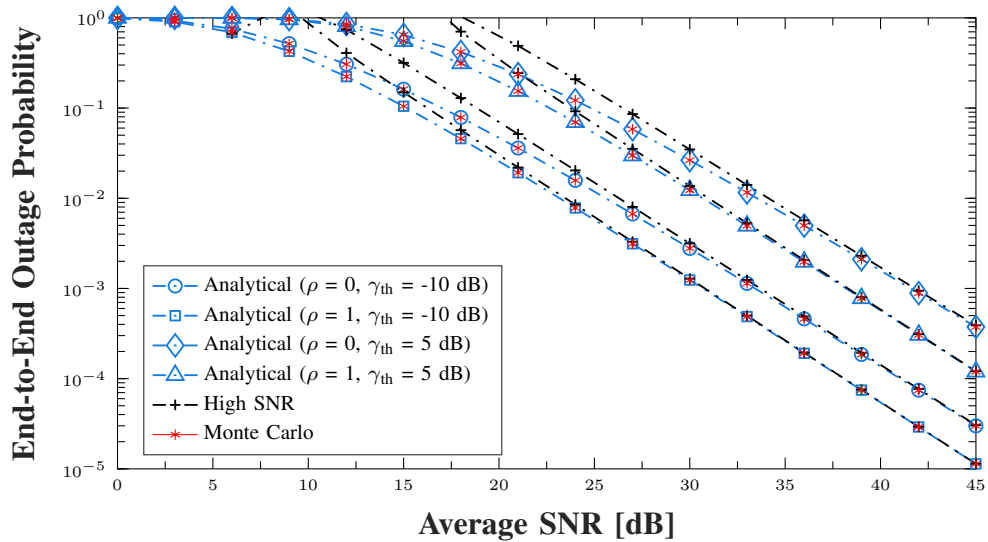
\begin{figure}[H]
\centering
\setlength\fheight{6cm}
\setlength\fwidth{12cm}
%
%
\definecolor{mycolor1}{rgb}{0.07059,0.48235,0.88627}%
\begin{tikzpicture}

\begin{axis}[%
width=0.951\fwidth,
height=\fheight,
at={(0\fwidth,0\fheight)},
scale only axis,
xmin=0,
xmax=45,
xlabel style={font=\bfseries\color{white!15!black}},
xlabel={Average SNR [dB]},
ymode=log,
ymin=1e-05,
ymax=1,
yminorticks=true,
ylabel style={font=\bfseries\color{white!15!black}},
ylabel={End-to-End Outage Probability},
axis background/.style={fill=white},
legend style={at={(0.03,0.03)}, anchor=south west, legend cell align=left, align=left, draw=white!15!black}
]
\addplot [color=mycolor1, dash pattern={on 5pt off 3pt on 1pt off 3pt} , line width=0.7pt, mark size=2.5pt, mark=o, mark options={solid, mycolor1}]
  table[row sep=crcr]{%
0	0.993313478489403\\
3	0.931001663518904\\
6	0.756189071819879\\
9	0.517945532062672\\
12	0.306358195209073\\
15	0.161905380789418\\
18	0.0787220256553657\\
21	0.0360194745186644\\
24	0.0157733048594673\\
27	0.00669422442593204\\
30	0.00277918593692317\\
33	0.00113659060143023\\
36	0.000460320195879715\\
39	0.000185384255611987\\
42	7.448645939101e-05\\
45	2.99415710330875e-05\\
};
\addlegendentry{$\text{Analytical (}\rho\text{ = 0, }\gamma{}_{\text{th}}\text{ = -10 dB)}$}

\addplot [color=mycolor1, dash pattern={on 5pt off 3pt on 1pt off 3pt} , line width=0.7pt, mark size=1.8pt, mark=square, mark options={solid, mycolor1}]
  table[row sep=crcr]{%
0	0.990625861615964\\
3	0.906912680379405\\
6	0.689525355362951\\
9	0.425794131786563\\
12	0.223254537697565\\
15	0.104784881637589\\
18	0.0458318599361032\\
21	0.019184500930476\\
24	0.00781402336207315\\
27	0.00312889815948103\\
30	0.00123949656959343\\
33	0.000487688363207006\\
36	0.000191052694733629\\
39	7.46371800876e-05\\
42	2.91058514826359e-05\\
45	1.1337176335724e-05\\
};
\addlegendentry{$\text{Analytical (}\rho\text{ = 1, }\gamma{}_{\text{th}}\text{ = -10 dB)}$}

\addplot [color=mycolor1, dash pattern={on 5pt off 3pt on 1pt off 3pt} , line width=0.7pt, mark size=4.3pt, mark=diamond, mark options={solid, mycolor1}]
  table[row sep=crcr]{%
0	0.999999993044397\\
3	0.999977336645213\\
6	0.997775127806117\\
9	0.968724448783632\\
12	0.854741027070394\\
15	0.647633316594461\\
18	0.419185288131079\\
21	0.237755952283483\\
24	0.121988012297722\\
27	0.0582198650379168\\
30	0.0264161115436676\\
33	0.011583709637127\\
36	0.00497002565411908\\
39	0.00210602382534286\\
42	0.000887780849136655\\
45	0.000374427065664484\\
};
\addlegendentry{$\text{Analytical (}\rho\text{ = 0, }\gamma{}_{\text{th}}\text{ = 5 dB)}$}

\addplot [color=mycolor1, dash pattern={on 5pt off 3pt on 1pt off 3pt} , line width=0.7pt, mark size=3.7pt, mark=triangle, mark options={solid, mycolor1}]
  table[row sep=crcr]{%
0	0.999999997432974\\
3	0.999985868830987\\
6	0.997828368319639\\
9	0.961112410540162\\
12	0.808271349543146\\
15	0.552859070919401\\
18	0.312940825477709\\
21	0.154485374710111\\
24	0.0697461943518316\\
27	0.0297729084559007\\
30	0.0122750764315998\\
33	0.0049526410025571\\
36	0.00197136181568147\\
39	0.000778001439934894\\
42	0.000305373801483988\\
45	0.00011944629106464\\
};
\addlegendentry{$\text{Analytical (}\rho\text{ = 1, }\gamma{}_{\text{th}}\text{ = 5 dB)}$}

\addplot [color=black, dash pattern={on 5pt off 3pt on 1pt off 3pt} , line width=0.7pt, mark=+, mark options={solid, black}]
  table[row sep=crcr]{%
0	3185.50195065279\\
3	15.1483791354627\\
6	0.670356380334734\\
9	1.42838128968029\\
12	0.739893658672721\\
15	0.317185145278788\\
18	0.128997081902084\\
21	0.0514653279299646\\
24	0.0203628597371006\\
27	0.00802602252436191\\
30	0.00315880295951421\\
33	0.00124329321708883\\
36	0.000489976456253816\\
39	0.000193550932106601\\
42	7.67181050610466e-05\\
45	3.0547427590788e-05\\
};
\addlegendentry{High SNR}

\addplot [color=red, draw=none, mark size=2.0pt, mark=asterisk, mark options={solid, red}]
  table[row sep=crcr]{%
0	0.993313478489403\\
3	0.931001663518904\\
6	0.756189071819879\\
9	0.517945532062672\\
12	0.306358195209073\\
15	0.161905380789418\\
18	0.0787220256553657\\
21	0.0360194745186644\\
24	0.0157733048594673\\
27	0.00669422442593204\\
30	0.00277918593692317\\
33	0.00113659060143023\\
36	0.000460320195879715\\
39	0.000185384255611987\\
42	7.448645939101e-05\\
45	2.99415710330875e-05\\
};
\addlegendentry{Monte Carlo}

\addplot [color=black, dash pattern={on 5pt off 3pt on 1pt off 3pt} , line width=0.7pt, mark=+, mark options={solid, black}, forget plot]
  table[row sep=crcr]{%
0	-7518.26820759506\\
3	92.0654452903929\\
6	6.50402730442056\\
9	1.27279049315715\\
12	0.40841423249285\\
15	0.149499944478982\\
18	0.0569125100724586\\
21	0.0219527991539964\\
24	0.00850773180198094\\
27	0.00330295577351869\\
30	0.00128319433740298\\
33	0.000498661840925507\\
36	0.000193808745526991\\
39	7.53294233785118e-05\\
42	2.92797294405567e-05\\
45	1.13808517866465e-05\\
};
\addplot [color=black, dash pattern={on 5pt off 3pt on 1pt off 3pt} , line width=0.7pt, mark=+, mark options={solid, black}, forget plot]
  table[row sep=crcr]{%
0	14326970.9699618\\
3	145662.597691353\\
6	2958.13855353446\\
9	52.7293336361301\\
12	-0.48500734075935\\
15	1.51291144109464\\
18	1.03652625765496\\
21	0.489864927098573\\
24	0.208887096559242\\
27	0.0858656628368398\\
30	0.0348088234706992\\
33	0.0140539654740773\\
36	0.00568124121002089\\
39	0.00230742527627437\\
42	0.000944080727833252\\
45	0.00039000256156807\\
};
\addplot [color=black,dash pattern={on 5pt off 3pt on 1pt off 3pt} , line width=0.7pt, mark=+, mark options={solid, black}, forget plot]
  table[row sep=crcr]{%
0	-34129982.7644662\\
3	-3948903.0667349\\
6	14961.2503497965\\
9	306.925224022579\\
12	24.830515425163\\
15	2.7102133707477\\
18	0.703870858453762\\
21	0.245168091726891\\
24	0.0919848669112893\\
27	0.0353098098901883\\
30	0.0136607233658483\\
33	0.00530011192844737\\
36	0.00205857236624207\\
39	0.000799899139854232\\
42	0.000310873187273519\\
45	0.000120827540842194\\
};
\addplot [color=red, draw=none, mark size=2.0pt, mark=asterisk, mark options={solid, red}, forget plot]
  table[row sep=crcr]{%
0	0.990625861615964\\
3	0.906912680379405\\
6	0.689525355362951\\
9	0.425794131786563\\
12	0.223254537697565\\
15	0.104784881637589\\
18	0.0458318599361032\\
21	0.019184500930476\\
24	0.00781402336207315\\
27	0.00312889815948103\\
30	0.00123949656959343\\
33	0.000487688363207006\\
36	0.000191052694733629\\
39	7.46371800876e-05\\
42	2.91058514826359e-05\\
45	1.1337176335724e-05\\
};
\addplot [color=red, draw=none, mark size=2.0pt, mark=asterisk, mark options={solid, red}, forget plot]
  table[row sep=crcr]{%
0	0.999999993044397\\
3	0.999977336645213\\
6	0.997775127806117\\
9	0.968724448783632\\
12	0.854741027070394\\
15	0.647633316594461\\
18	0.419185288131079\\
21	0.237755952283483\\
24	0.121988012297722\\
27	0.0582198650379168\\
30	0.0264161115436676\\
33	0.011583709637127\\
36	0.00497002565411908\\
39	0.00210602382534286\\
42	0.000887780849136655\\
45	0.000374427065664484\\
};
\addplot [color=red, draw=none, mark size=2.0pt, mark=asterisk, mark options={solid, red}, forget plot]
  table[row sep=crcr]{%
0	0.999999997432974\\
3	0.999985868830987\\
6	0.997828368319639\\
9	0.961112410540162\\
12	0.808271349543146\\
15	0.552859070919401\\
18	0.312940825477709\\
21	0.154485374710111\\
24	0.0697461943518316\\
27	0.0297729084559007\\
30	0.0122750764315998\\
33	0.0049526410025571\\
36	0.00197136181568147\\
39	0.000778001439934894\\
42	0.000305373801483988\\
45	0.00011944629106464\\
};
\end{axis}
\end{tikzpicture}%
    \caption{Probability of outage for Fixed relaying gain. The SEL is issumed while IM/DD is the receive detection mode. Outdated and perfect CSIs are assumed with different SNDR thresholds.}
\end{figure}
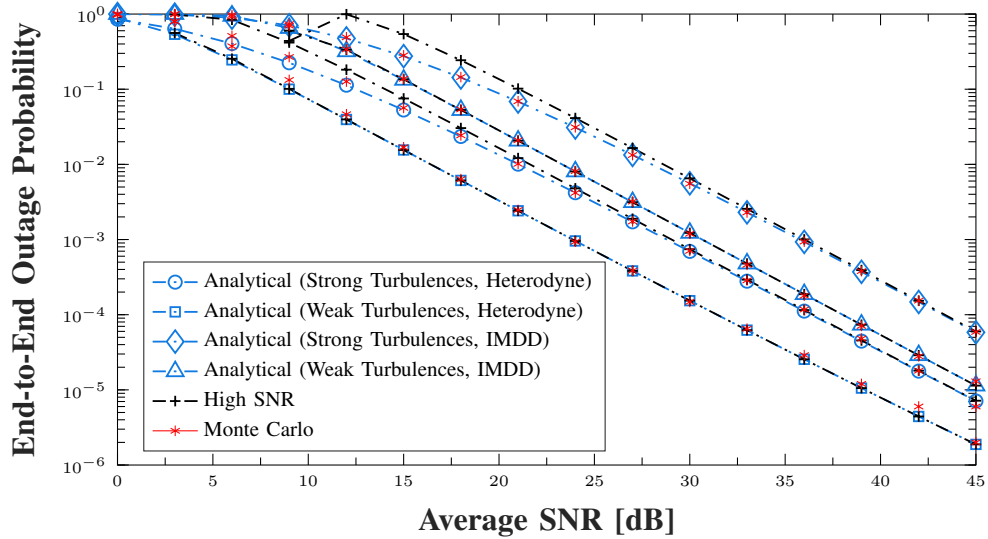
\begin{figure}[H]
\centering
\setlength\fheight{6cm}
\setlength\fwidth{12cm}
%
%
\definecolor{mycolor1}{rgb}{0.07059,0.48235,0.88627}%
\begin{tikzpicture}

\begin{axis}[%
width=0.951\fwidth,
height=\fheight,
at={(0\fwidth,0\fheight)},
scale only axis,
xmin=0,
xmax=45,
xlabel style={font=\bfseries\color{white!15!black}},
xlabel={Average SNR [dB]},
ymode=log,
ymin=1e-06,
ymax=1,
yminorticks=true,
ylabel style={font=\bfseries\color{white!15!black}},
ylabel={End-to-End Outage Probability},
axis background/.style={fill=white},
legend style={at={(0.03,0.03)}, anchor=south west, legend cell align=left, align=left, draw=white!15!black}
]
\addplot [color=mycolor1, dash pattern={on 5pt off 3pt on 1pt off 3pt} , line width=0.7pt, mark size=2.5pt, mark=o, mark options={solid, mycolor1}]
  table[row sep=crcr]{%
0	0.855093000040345\\
3	0.637639421781055\\
6	0.404435510409476\\
9	0.224806624570666\\
12	0.113116858296884\\
15	0.052934036100051\\
18	0.0235152485798567\\
21	0.010067173009643\\
24	0.00419902392981966\\
27	0.00171995488775261\\
30	0.000695896930558684\\
33	0.000279335091513639\\
36	0.000111612577604456\\
39	4.45109446398947e-05\\
42	1.77564207673839e-05\\
45	7.09962118730523e-06\\
};
\addlegendentry{Analytical (Strong Turbulences, Heterodyne)}

\addplot [color=mycolor1, dash pattern={on 5pt off 3pt on 1pt off 3pt} , line width=0.7pt, mark size=1.8pt, mark=square, mark options={solid, mycolor1}]
  table[row sep=crcr]{%
0	0.882779483475451\\
3	0.537918147601011\\
6	0.245694357302646\\
9	0.100140489595294\\
12	0.0395484834714863\\
15	0.0155358160709889\\
18	0.00611329914786124\\
21	0.00241458080407691\\
24	0.000958267566993959\\
27	0.000382512905345528\\
30	0.000153752450370753\\
33	6.23150234804316e-05\\
36	2.55033272684233e-05\\
39	1.05561018715615e-05\\
42	4.42557807361229e-06\\
45	1.88186137250364e-06\\
};
\addlegendentry{Analytical (Weak Turbulences, Heterodyne)}

\addplot [color=mycolor1, dash pattern={on 5pt off 3pt on 1pt off 3pt} , line width=0.7pt, mark size=4.3pt, mark=diamond, mark options={solid, mycolor1}]
  table[row sep=crcr]{%
0	0.999019629842702\\
3	0.981620789468226\\
6	0.894646497817439\\
9	0.706656527657042\\
12	0.474040593901837\\
15	0.275482383051871\\
18	0.143182550283271\\
21	0.0685384625041472\\
24	0.030916220333801\\
27	0.0133654256234675\\
30	0.00560547694629123\\
33	0.00230075130119071\\
36	0.000929996794292127\\
39	0.000371893667140787\\
42	0.00014761017859388\\
45	5.82947053923653e-05\\
};
\addlegendentry{Analytical (Strong Turbulences, IMDD)}

\addplot [color=mycolor1, dash pattern={on 5pt off 3pt on 1pt off 3pt} , line width=0.7pt, mark size=3.7pt, mark=triangle, mark options={solid, mycolor1}]
  table[row sep=crcr]{%
0	0.999999999086601\\
3	0.999613607340667\\
6	0.944922574290965\\
9	0.651325102425812\\
12	0.319147739451114\\
15	0.13316182051833\\
18	0.0527616477082107\\
21	0.020630674430158\\
24	0.00804605006906034\\
27	0.00313815656694862\\
30	0.00122492912913275\\
33	0.000478687195488181\\
36	0.000187347643264568\\
39	7.3464921127348e-05\\
42	2.8878170980472e-05\\
45	1.13865518155203e-05\\
};
\addlegendentry{Analytical (Weak Turbulences, IMDD)}

\addplot [color=black, dash pattern={on 5pt off 3pt on 1pt off 3pt} , line width=0.7pt, mark=+, mark options={solid, black}]
  table[row sep=crcr]{%
0	-3.67226383385774\\
3	0.985013066022023\\
6	0.840289957574856\\
9	0.417403958901935\\
12	0.181891560639008\\
15	0.0752874808633094\\
18	0.0304015205668632\\
21	0.0121158595802241\\
24	0.00479369800310987\\
27	0.00188943675022582\\
30	0.000743518646105367\\
33	0.00029256547744299\\
36	0.000115254481933192\\
39	4.55057717849262e-05\\
42	1.80264104282496e-05\\
45	7.17248847337233e-06\\
};
\addlegendentry{High SNR}

\addplot [color=red, draw=none, mark size=2.0pt, mark=asterisk, mark options={solid, red}]
  table[row sep=crcr]{%
0	0.971598\\
3	0.802218\\
6	0.511477\\
9	0.268398\\
12	0.126498\\
15	0.056487\\
18	0.024315\\
21	0.010154\\
24	0.004178\\
27	0.001742\\
30	0.000707\\
33	0.000286\\
36	0.000117\\
39	4.8e-05\\
42	1.8e-05\\
45	6e-06\\
};
\addlegendentry{Monte Carlo}

\addplot [color=black, dash pattern={on 5pt off 3pt on 1pt off 3pt} , line width=0.7pt, mark=+, mark options={solid, black}, forget plot]
  table[row sep=crcr]{%
0	-0.73096716157488\\
3	0.560259693297475\\
6	0.254206809886553\\
9	0.10111989336732\\
12	0.0396381049057516\\
15	0.0155432237295592\\
18	0.00611387703783974\\
21	0.00241462423447592\\
24	0.000958270746767096\\
27	0.000382513133700501\\
30	0.000153752466528082\\
33	6.23150246102433e-05\\
36	2.55033273466721e-05\\
39	1.0556101876938e-05\\
42	4.42557807397924e-06\\
45	1.88186137252854e-06\\
};
\addplot [color=red, draw=none, mark size=2.0pt, mark=asterisk, mark options={solid, red}, forget plot]
  table[row sep=crcr]{%
0	0.993483\\
3	0.793563\\
6	0.372727\\
9	0.13279\\
12	0.046462\\
15	0.01687\\
18	0.006361\\
21	0.002474\\
24	0.000935\\
27	0.000385\\
30	0.000148\\
33	6.5e-05\\
36	2.9e-05\\
39	1.2e-05\\
42	6e-06\\
45	2e-06\\
};
\addplot [color=black, dash pattern={on 5pt off 3pt on 1pt off 3pt} , line width=0.7pt, mark=+, mark options={solid, black}, forget plot]
  table[row sep=crcr]{%
0	-804.98963096851\\
3	-110.327541168239\\
6	-11.4352186541322\\
9	0.438794657912117\\
12	0.99429523167329\\
15	0.541727645449184\\
18	0.243039843515648\\
21	0.101815478760916\\
24	0.0413259583097554\\
27	0.0164948138581308\\
30	0.00652075133881988\\
33	0.00256310890990976\\
36	0.00100404894554174\\
39	0.000392542089956831\\
42	0.000153311096144349\\
45	5.98558886632614e-05\\
};
\addplot [color=black, dash pattern={on 5pt off 3pt on 1pt off 3pt} , line width=0.7pt, mark=+, mark options={solid, black}, forget plot]
  table[row sep=crcr]{%
0	-20656.5430303697\\
3	-385.324446460911\\
6	-6.48765801219217\\
9	0.601819107067616\\
12	0.335196424124843\\
15	0.135326001452442\\
18	0.0529704956677242\\
21	0.0206484111683308\\
24	0.00804745681389371\\
27	0.00313826347315278\\
30	0.0012249370193533\\
33	0.000478687765562193\\
36	0.000187347683791384\\
39	7.34649239719761e-05\\
42	2.88781711780995e-05\\
45	1.13865518291345e-05\\
};
\addplot [color=red, draw=none, mark size=2.0pt, mark=asterisk, mark options={solid, red}, forget plot]
  table[row sep=crcr]{%
0	0.999887\\
3	0.992187\\
6	0.92308\\
9	0.735122\\
12	0.49046\\
15	0.282206\\
18	0.145126\\
21	0.069115\\
24	0.031166\\
27	0.01331\\
30	0.005529\\
33	0.00229\\
36	0.000938\\
39	0.00038\\
42	0.000154\\
45	5.9e-05\\
};
\addplot [color=red, draw=none, mark size=2.0pt, mark=asterisk, mark options={solid, red}, forget plot]
  table[row sep=crcr]{%
0	1\\
3	0.999971\\
6	0.971307\\
9	0.699083\\
12	0.338828\\
15	0.137712\\
18	0.053898\\
21	0.020906\\
24	0.008129\\
27	0.003198\\
30	0.001194\\
33	0.000461\\
36	0.000179\\
39	7.1e-05\\
42	2.8e-05\\
45	1.3e-05\\
};
\end{axis}
\end{tikzpicture}%
    \caption{Probability of outage for variable relaying gain and TWTA impairment. This scenario considers strong and severe atmospheric turbulences with heterodyne and IMDD receive detection techniques.}
\end{figure}
\vspace*{-1cm}
Fig.~3 shows the dependence of the OP of FG relaying with respect to the average SNR considering various values of the outage threshold $\gamma_{\text{th}}$ and the time correlation coefficient $\rho$. In addition, the relays are supposed to be impaired by SEL impairments and the receiver employs the IM/DD as a method of detection. For both correlation values, we observe that the performance deteriorates as the $\gamma_{\text{th}}$ becomes higher and this result is certainly expected since for a given SNDR, the probability that the SNDR falls below a higher outage threshold becomes higher. For a given threshold, the system works better when the best relay of the last rank ($m = N$) is selected according to PRS protocol. We observe that the performance improves as the correlation coefficient increases. For a perfect CSI estimation ($\rho = 1$), there are full correlation between the two CSIs and the selection of the best relay is certainly achieved based on the feedback or the outdated CSI. However, for a completely outdated CSI ($\rho = 0$) the two CSIs are completely uncorrelated and hence the selection of the best relay is uncertain since the selection is based on a completely outdated CSI. As a result, the performance deteriorates substantially.\\
Fig.~4 illustrates the variations of the OP of VG relaying versus the average SNR for moderate and strong atmospheric turbulences considering both the heterodyne and IM/DD as a detection scheme at the receiver. For moderate turblence (higher values of $\alpha_1$, $\alpha_2$), the system works better for the heterodyne mode compared to IM/DD. As the turbulence-induced fading becomes severe (lower values of $\alpha_1$, $\alpha_2$), the system performs worse compared to the first case. We also observe that the system works better for IM/DD under moderate turbulence than the heterodyne mode for severe turbulences even though the heterodyne mode outperforms the IM/DD. It turned out that the system depends to a large extent on the state of the optical channel.\\
Fig.~5 provides the variations of the OP for FG relaying against the average SNR for different values of the pointing error coefficients. In addition, the relays suffer from TWTA impairments and the receiver detects the incoming signal using IM/DD method. We observe that the system works better as the pointing error coefficient decreases. In fact, as this coefficient $\xi$ decreases, the pointing error fading becomes more severe. For a given average SNR of 30 dB, the system achieves roughly the following outage values 2~10$^{-3}$, 2.5~10$^{-2}$, 0.1 and 0.5 for the pointing error coefficients equal to 0.4, 0.7, 0.9 and 1.2, respectively. It turned out that the outage performance gets better as the pointing error coefficient becomes higher and thereby we prove again that the system depends substantially on the state of the optical channel.
\vspace*{-0.5cm}
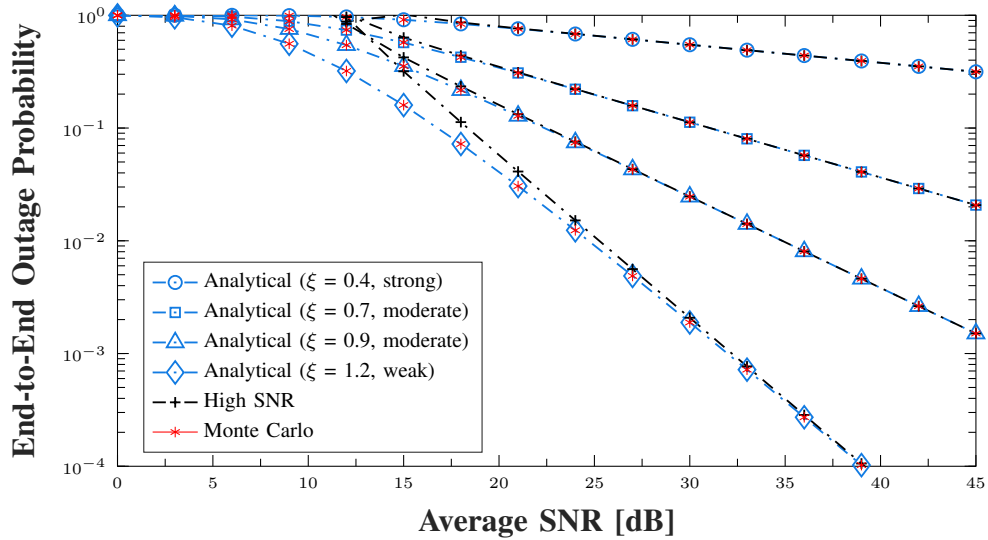
\begin{figure}[H]
\centering
\setlength\fheight{6cm}
\setlength\fwidth{12cm}
%
%
\definecolor{mycolor1}{rgb}{0.07059,0.48235,0.88627}%
\begin{tikzpicture}

\begin{axis}[%
width=0.951\fwidth,
height=\fheight,
at={(0\fwidth,0\fheight)},
scale only axis,
xmin=0,
xmax=45,
xlabel style={font=\bfseries\color{white!15!black}},
xlabel={Average SNR [dB]},
ymode=log,
ymin=0.0001,
ymax=1,
yminorticks=true,
ylabel style={font=\bfseries\color{white!15!black}},
ylabel={End-to-End Outage Probability},
axis background/.style={fill=white},
legend style={at={(0.03,0.03)}, anchor=south west, legend cell align=left, align=left, draw=white!15!black}
]
\addplot [color=mycolor1, dash pattern={on 5pt off 3pt on 1pt off 3pt}, line width=0.7pt, mark size=2.5pt, mark=o, mark options={solid, mycolor1}]
  table[row sep=crcr]{%
0	0.999999994261501\\
3	0.999992924342518\\
6	0.999494987404986\\
9	0.993269335623046\\
12	0.967242582618342\\
15	0.913081124352278\\
18	0.83923653079036\\
21	0.759493978302906\\
24	0.682598278310203\\
27	0.611910946830633\\
30	0.548076711256951\\
33	0.490774558253843\\
36	0.439431405318964\\
39	0.393452345138288\\
42	0.352282924499218\\
45	0.315421278502505\\
};
\addlegendentry{$\text{Analytical (}\xi\text{ = 0.4, strong)}$}

\addplot [color=mycolor1, dash pattern={on 5pt off 3pt on 1pt off 3pt}, line width=0.7pt, mark size=1.8pt, mark=square, mark options={solid, mycolor1}]
  table[row sep=crcr]{%
0	0.999967741233785\\
3	0.997687967530225\\
6	0.972955512049497\\
9	0.887194629780678\\
12	0.739806344112882\\
15	0.57331087902648\\
18	0.424871931025519\\
21	0.307864471646247\\
24	0.220872665334535\\
27	0.157824468206613\\
30	0.112600152526521\\
33	0.0802897664446183\\
36	0.0572395818017738\\
39	0.0408041864776296\\
42	0.0290873858843698\\
45	0.0207349400678551\\
};
\addlegendentry{$\text{Analytical (}\xi\text{ = 0.7, moderate)}$}

\addplot [color=mycolor1, dash pattern={on 5pt off 3pt on 1pt off 3pt}, line width=0.7pt, mark size=3.7pt, mark=triangle, mark options={solid, mycolor1}]
  table[row sep=crcr]{%
0	0.999668018073326\\
3	0.98884606082321\\
6	0.919566509437539\\
9	0.755467422676878\\
12	0.543859144878596\\
15	0.35433736118161\\
18	0.216920696855599\\
21	0.128282733757047\\
24	0.0745254013977797\\
27	0.0429189508582617\\
30	0.0246144636668759\\
33	0.0140893157918331\\
36	0.00805752152375261\\
39	0.00460612815022599\\
42	0.00263263483825904\\
45	0.00150456054418857\\
};
\addlegendentry{$\text{Analytical (}\xi\text{ = 0.9, moderate)}$}

\addplot [color=mycolor1,dash pattern={on 5pt off 3pt on 1pt off 3pt}, line width=0.7pt, mark size=4.3pt, mark=diamond, mark options={solid, mycolor1}]
  table[row sep=crcr]{%
0	0.997831517740287\\
3	0.960478658664862\\
6	0.809316988180463\\
9	0.5590764270882\\
12	0.32075989989361\\
15	0.159834914832008\\
18	0.0722196640265641\\
21	0.0305635175056875\\
24	0.0123866913507179\\
27	0.00487835421942229\\
30	0.00188504314272553\\
33	0.000719200137789366\\
36	0.000272090520345825\\
39	0.000102376578630325\\
42	3.83922538979009e-05\\
45	1.43732906823235e-05\\
};
\addlegendentry{$\text{Analytical (}\xi\text{ = 1.2, weak)}$}

\addplot [color=black, dash pattern={on 5pt off 3pt on 1pt off 3pt}, line width=0.7pt, mark=+, mark options={solid, black}]
  table[row sep=crcr]{%
0	572218403.128299\\
3	1693163.30481773\\
6	3961.33212910711\\
9	-12.8838134618543\\
12	0.880308482131184\\
15	1.01641490781489\\
18	0.863318673953944\\
21	0.765162946802328\\
24	0.684030248297242\\
27	0.612284384910765\\
30	0.548175033242717\\
33	0.490800480969237\\
36	0.439438232179029\\
39	0.393454139843282\\
42	0.352283395434329\\
45	0.315421401857819\\
};
\addlegendentry{High SNR}

\addplot [color=red, draw=none, mark size=2.0pt, mark=asterisk, mark options={solid, red}]
  table[row sep=crcr]{%
0	0.999999994261501\\
3	0.999992924342518\\
6	0.999494987404986\\
9	0.993269335623046\\
12	0.967242582618342\\
15	0.913081124352278\\
18	0.83923653079036\\
21	0.759493978302906\\
24	0.682598278310203\\
27	0.611910946830633\\
30	0.548076711256951\\
33	0.490774558253843\\
36	0.439431405318964\\
39	0.393452345138288\\
42	0.352282924499218\\
45	0.315421278502505\\
};
\addlegendentry{Monte Carlo}

\addplot [color=black, dash pattern={on 5pt off 3pt on 1pt off 3pt}, line width=0.7pt, mark=+, mark options={solid, black}, forget plot]
  table[row sep=crcr]{%
0	1575562.21752037\\
3	2100.24432530111\\
6	-36.6037993665053\\
9	1.27933400653437\\
12	1.00923631314309\\
15	0.635142145738567\\
18	0.43980561952012\\
21	0.311682297691243\\
24	0.221871130396438\\
27	0.158087264647394\\
30	0.112669364015353\\
33	0.0803079731306771\\
36	0.0572443633136573\\
39	0.040805440046369\\
42	0.029087713981898\\
45	0.0207350258051715\\
};
\addplot [color=black, dash pattern={on 5pt off 3pt on 1pt off 3pt}, line width=0.7pt, mark=+, mark options={solid, black}, forget plot]
  table[row sep=crcr]{%
0	185515.377654445\\
3	-416.712531441442\\
6	-10.8946144989263\\
9	1.85481022728643\\
12	0.838030715374541\\
15	0.423213811992657\\
18	0.234142702577431\\
21	0.13274599517924\\
24	0.0756963961948651\\
27	0.043227062740844\\
30	0.0246955084677259\\
33	0.0141106053511116\\
36	0.00806310509430619\\
39	0.00460759015118162\\
42	0.00263301704425778\\
45	0.0015046603137594\\
};
\addplot [color=black, dash pattern={on 5pt off 3pt on 1pt off 3pt}, line width=0.7pt, mark=+, mark options={solid, black}, forget plot]
  table[row sep=crcr]{%
0	16430.8844417557\\
3	-392.926970216123\\
6	4.76988507068508\\
9	3.25130400100159\\
12	0.964628379056005\\
15	0.318450113479019\\
18	0.112854820911768\\
21	0.0411488065148462\\
24	0.0151593300298088\\
27	0.0056054208126316\\
30	0.0020756030303628\\
33	0.000769088447946098\\
36	0.000285133729427733\\
39	0.000105781987362885\\
42	3.92801562218992e-05\\
45	1.46044958843072e-05\\
};
\addplot [color=red, draw=none, mark size=2.0pt, mark=asterisk, mark options={solid, red}, forget plot]
  table[row sep=crcr]{%
0	0.999967741233785\\
3	0.997687967530225\\
6	0.972955512049497\\
9	0.887194629780678\\
12	0.739806344112882\\
15	0.57331087902648\\
18	0.424871931025519\\
21	0.307864471646247\\
24	0.220872665334535\\
27	0.157824468206613\\
30	0.112600152526521\\
33	0.0802897664446183\\
36	0.0572395818017738\\
39	0.0408041864776296\\
42	0.0290873858843698\\
45	0.0207349400678551\\
};
\addplot [color=red, draw=none, mark size=2.0pt, mark=asterisk, mark options={solid, red}, forget plot]
  table[row sep=crcr]{%
0	0.999668018073326\\
3	0.98884606082321\\
6	0.919566509437539\\
9	0.755467422676878\\
12	0.543859144878596\\
15	0.35433736118161\\
18	0.216920696855599\\
21	0.128282733757047\\
24	0.0745254013977797\\
27	0.0429189508582617\\
30	0.0246144636668759\\
33	0.0140893157918331\\
36	0.00805752152375261\\
39	0.00460612815022599\\
42	0.00263263483825904\\
45	0.00150456054418857\\
};
\addplot [color=red, draw=none, mark size=2.0pt, mark=asterisk, mark options={solid, red}, forget plot]
  table[row sep=crcr]{%
0	0.997831517740287\\
3	0.960478658664862\\
6	0.809316988180463\\
9	0.5590764270882\\
12	0.32075989989361\\
15	0.159834914832008\\
18	0.0722196640265641\\
21	0.0305635175056875\\
24	0.0123866913507179\\
27	0.00487835421942229\\
30	0.00188504314272553\\
33	0.000719200137789366\\
36	0.000272090520345825\\
39	0.000102376578630325\\
42	3.83922538979009e-05\\
45	1.43732906823235e-05\\
};
\end{axis}
\end{tikzpicture}%
    \caption{Probability of outage for variable relaying gain with IM/DD receive detection and TWTA impairment model. Simulation is based upon different pointing errors severities.}
\end{figure}
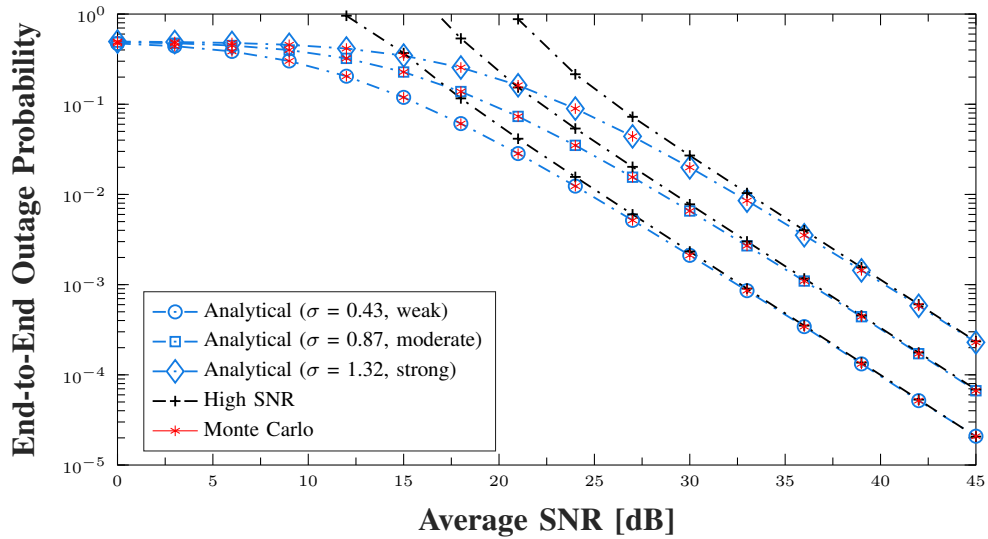
\begin{figure}[H]
\centering
\setlength\fheight{6cm}
\setlength\fwidth{12cm}
%
%
\definecolor{mycolor1}{rgb}{0.07059,0.48235,0.88627}%
\begin{tikzpicture}

\begin{axis}[%
width=0.951\fwidth,
height=\fheight,
at={(0\fwidth,0\fheight)},
scale only axis,
xmin=0,
xmax=45,
xlabel style={font=\bfseries\color{white!15!black}},
xlabel={Average SNR [dB]},
ymode=log,
ymin=1e-05,
ymax=1,
yminorticks=true,
ylabel style={font=\bfseries\color{white!15!black}},
ylabel={End-to-End Outage Probability},
axis background/.style={fill=white},
legend style={at={(0.03,0.03)}, anchor=south west, legend cell align=left, align=left, draw=white!15!black}
]
\addplot [color=mycolor1, dash pattern={on 5pt off 3pt on 1pt off 3pt} , line width=0.7pt, mark size=2.5pt, mark=o, mark options={solid, mycolor1}]
  table[row sep=crcr]{%
0	0.47205938778072\\
3	0.440932236803358\\
6	0.38527070360884\\
9	0.301839857014136\\
12	0.204613793094622\\
15	0.119201837980071\\
18	0.061023738371236\\
21	0.0283971960870695\\
24	0.0123682515327834\\
27	0.00516135891661451\\
30	0.00211174902326956\\
33	0.000857293714388682\\
36	0.000342447689170813\\
39	0.00013220710123813\\
42	5.17553460984458e-05\\
45	2.08723445702807e-05\\
};
\addlegendentry{$\text{Analytical (}\sigma\text{ = 0.43, weak)}$}

\addplot [color=mycolor1, dash pattern={on 5pt off 3pt on 1pt off 3pt} , line width=0.7pt, mark size=1.8pt, mark=square, mark options={solid, mycolor1}]
  table[row sep=crcr]{%
0	0.488281922588352\\
3	0.474664876341238\\
6	0.4478905033772\\
9	0.399140778052537\\
12	0.32267472856433\\
15	0.227697679290949\\
18	0.138238084714684\\
21	0.073304000300459\\
24	0.0350202188411734\\
27	0.0155366105796513\\
30	0.00655499876661983\\
33	0.0026948504534552\\
36	0.00109648365526102\\
39	0.000441207145618865\\
42	0.000171607366433189\\
45	6.66417722184382e-05\\
};
\addlegendentry{$\text{Analytical (}\sigma\text{ = 0.87, moderate)}$}

\addplot [color=mycolor1, dash pattern={on 5pt off 3pt on 1pt off 3pt} , line width=0.7pt, mark size=4.3pt, mark=diamond, mark options={solid, mycolor1}]
  table[row sep=crcr]{%
0	0.495226068600388\\
3	0.489629120371819\\
6	0.478335565482645\\
9	0.456154656887021\\
12	0.414834053495748\\
15	0.346641111250599\\
18	0.255465521345303\\
21	0.162300459864617\\
24	0.089507639116967\\
27	0.0440442140941846\\
30	0.0199534340391908\\
33	0.00853332730919803\\
36	0.0035295058936773\\
39	0.00143963501173885\\
42	0.000582390684904135\\
45	0.000229165125893665\\
};
\addlegendentry{$\text{Analytical (}\sigma\text{ = 1.32, strong)}$}

\addplot [color=black, dash pattern={on 5pt off 3pt on 1pt off 3pt} , line width=0.7pt, mark=+, mark options={solid, black}]
  table[row sep=crcr]{%
0	-660433.490777754\\
3	-36628.0941180777\\
6	-874.283994195661\\
9	-20.7131093654223\\
12	0.95730229255612\\
15	0.369343162902463\\
18	0.116111808557817\\
21	0.0414702633267253\\
24	0.0156434754462568\\
27	0.00601481491958988\\
30	0.00232858401538838\\
33	0.000903857742765968\\
36	0.000351222094226611\\
39	0.000136549789952922\\
42	5.31054929935326e-05\\
45	2.06587042725914e-05\\
};
\addlegendentry{High SNR}

\addplot [color=red, draw=none, mark size=2.0pt, mark=asterisk, mark options={solid, red}]
  table[row sep=crcr]{%
0	0.47205938778072\\
3	0.440932236803358\\
6	0.38527070360884\\
9	0.301839857014136\\
12	0.204613793094622\\
15	0.119201837980071\\
18	0.061023738371236\\
21	0.0283971960870695\\
24	0.0123682515327834\\
27	0.00516135891661451\\
30	0.00211174902326956\\
33	0.000857293714388682\\
36	0.000342447689170813\\
39	0.00013220710123813\\
42	5.17553460984458e-05\\
45	2.08723445702807e-05\\
};
\addlegendentry{Monte Carlo}

\addplot [color=black, dash pattern={on 5pt off 3pt on 1pt off 3pt} , line width=0.7pt, mark=+, mark options={solid, black}, forget plot]
  table[row sep=crcr]{%
0	-835391187.973627\\
3	-37933322.2520301\\
6	-341394.663981064\\
9	112.309088086218\\
12	4.88424074588403\\
15	2.43503847376158\\
18	0.536156992200438\\
21	0.153428793074804\\
24	0.0538069184739474\\
27	0.0202312444686257\\
30	0.00777737665424927\\
33	0.00301180131515028\\
36	0.00116926357350816\\
39	0.000454350297810768\\
42	0.000176612927258679\\
45	6.86636554242392e-05\\
};
\addplot [color=black, dash pattern={on 5pt off 3pt on 1pt off 3pt} , line width=0.7pt, mark=+, mark options={solid, black}, forget plot]
  table[row sep=crcr]{%
0	-1125940155495.02\\
3	-50632798134.1946\\
6	-401848472.858591\\
9	2948826.47815207\\
12	47620.5408974783\\
15	285.848745349389\\
18	7.15653308404733\\
21	0.877363764482813\\
24	0.215116928687991\\
27	0.072571649393842\\
30	0.0270146104639541\\
33	0.0103575442222688\\
36	0.00400836491818157\\
39	0.00155593513743613\\
42	0.000604569714945991\\
45	0.0002349874092476\\
};
\addplot [color=red, draw=none, mark size=2.0pt, mark=asterisk, mark options={solid, red}, forget plot]
  table[row sep=crcr]{%
0	0.488281922588352\\
3	0.474664876341238\\
6	0.4478905033772\\
9	0.399140778052537\\
12	0.32267472856433\\
15	0.227697679290949\\
18	0.138238084714684\\
21	0.073304000300459\\
24	0.0350202188411734\\
27	0.0155366105796513\\
30	0.00655499876661983\\
33	0.0026948504534552\\
36	0.00109648365526102\\
39	0.000441207145618865\\
42	0.000171607366433189\\
45	6.66417722184382e-05\\
};
\addplot [color=red, draw=none, mark size=2.0pt, mark=asterisk, mark options={solid, red}, forget plot]
  table[row sep=crcr]{%
0	0.495226068600388\\
3	0.489629120371819\\
6	0.478335565482645\\
9	0.456154656887021\\
12	0.414834053495748\\
15	0.346641111250599\\
18	0.255465521345303\\
21	0.162300459864617\\
24	0.089507639116967\\
27	0.0440442140941846\\
30	0.0199534340391908\\
33	0.00853332730919803\\
36	0.0035295058936773\\
39	0.00143963501173885\\
42	0.000582390684904135\\
45	0.000229165125893665\\
};
\end{axis}
\end{tikzpicture}%
    \caption{Probability of outage for fixed relaying gain with IM/DD receive detection technique and SEL impairment. The scenario considers different weather attenuation conditions.}
\end{figure}
\vspace*{-1cm}
Fig.~6 presents the dependence of the outage probability of FG relaying versus the average SNR for various weather conditions. The system is also assumed to suffer from the SEL impairments while the IM/DD is adopted as the detection technique. We observe that for lower weather attenuation, the system works better. However, as the path loss becomes more severe, the performance gets worse. Thereby, the system proves its high dependence on the third component of the optical fading which is the atmospheric path loss.\\
Fig.~7 illustrates the variations of the BEP of FG relaying with respect to the average SNR for various modulation schemes given in table II. The relays are impaired by the SEL imperfection and the receiver uses the heterodyne mode to detect the incoming FSO signal. We observe the accuracy of the expression of BEP since it matches the exact Monte Carlo simulation. We also note that the system works better for BPSK, however, the performance gets much worse for 64-QAM modulation. In fact, there is a tradeoff between these two modulation schemes: BPSK yields lower error while the 64-QAM provides much more bandwidth efficiency which is very advantageous.
\vspace*{-0.5cm}
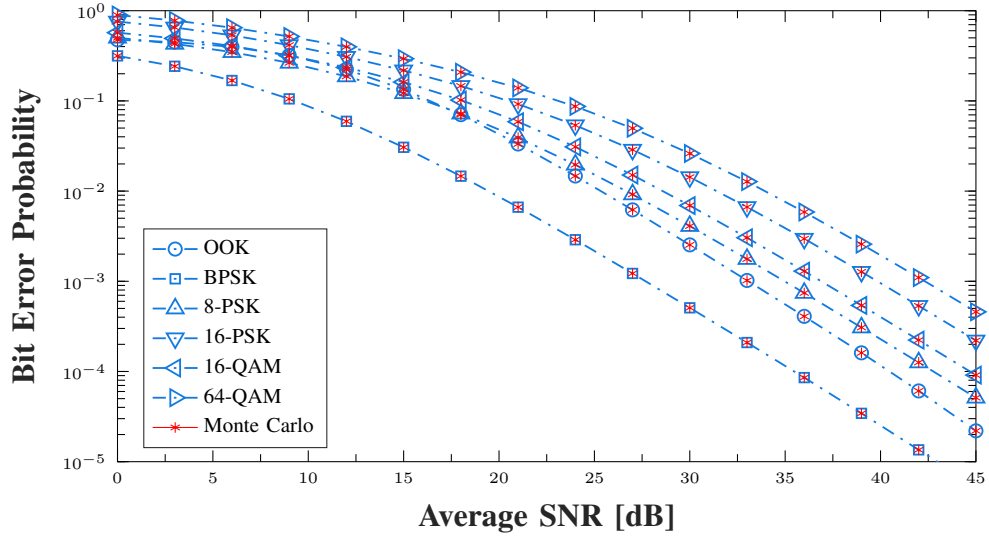
\begin{figure}[H]
\centering
\setlength\fheight{6cm}
\setlength\fwidth{12cm}
%
%
\definecolor{mycolor1}{rgb}{0.07059,0.48235,0.88627}%
\begin{tikzpicture}

\begin{axis}[%
width=0.951\fwidth,
height=\fheight,
at={(0\fwidth,0\fheight)},
scale only axis,
xmin=0,
xmax=45,
xlabel style={font=\bfseries\color{white!15!black}},
xlabel={Average SNR [dB]},
ymode=log,
ymin=1e-05,
ymax=1,
yminorticks=true,
ylabel style={font=\bfseries\color{white!15!black}},
ylabel={Bit Error Probability},
axis background/.style={fill=white},
legend style={at={(0.03,0.03)}, anchor=south west, legend cell align=left, align=left, draw=white!15!black}
]
\addplot [color=mycolor1, dash pattern={on 5pt off 3pt on 1pt off 3pt} , line width=0.7pt, mark size=2.5pt, mark=o, mark options={solid, mycolor1}]
  table[row sep=crcr]{%
0	0.475619623250311\\
3	0.448158110686024\\
6	0.397944542549645\\
9	0.319623410047207\\
12	0.223434979335542\\
15	0.134152689241089\\
18	0.0703366131498683\\
21	0.0332829024173976\\
24	0.0146965424410429\\
27	0.00619349902691856\\
30	0.00253374014717559\\
33	0.00102146313851809\\
36	0.000408570181131701\\
39	0.000160966476870366\\
42	6.08809739394978e-05\\
45	2.19660424578518e-05\\
};
\addlegendentry{OOK}

\addplot [color=mycolor1, dash pattern={on 5pt off 3pt on 1pt off 3pt} , line width=0.7pt, mark size=1.8pt, mark=square, mark options={solid, mycolor1}]
  table[row sep=crcr]{%
0	0.315768589748116\\
3	0.242069529758724\\
6	0.168328728428027\\
9	0.105299433137217\\
12	0.0594041684226517\\
15	0.0305962773614037\\
18	0.0146460039683937\\
21	0.00663568424441392\\
24	0.0028882316514401\\
27	0.00122180295185887\\
30	0.000507978570083546\\
33	0.000209403950144642\\
36	8.55360414775296e-05\\
39	3.43501664794626e-05\\
42	1.35251015939023e-05\\
45	5.3664090745105e-06\\
};
\addlegendentry{BPSK}

\addplot [color=mycolor1, dash pattern={on 5pt off 3pt on 1pt off 3pt} , line width=0.7pt, mark size=3.7pt, mark=triangle, mark options={solid, mycolor1}]
  table[row sep=crcr]{%
0	0.501708060665967\\
3	0.429071803615067\\
6	0.347534855460829\\
9	0.264324243123336\\
12	0.186962440895644\\
15	0.121655901075286\\
18	0.0722559602434827\\
21	0.03919415991085\\
24	0.0196083166109786\\
27	0.00918953636980934\\
30	0.00410114961759739\\
33	0.00176671856407097\\
36	0.000742397090255414\\
39	0.000307253191689341\\
42	0.000126016320745133\\
45	5.11876023008889e-05\\
};
\addlegendentry{8-PSK}

\addplot [color=mycolor1,dash pattern={on 5pt off 3pt on 1pt off 3pt} , line width=0.7pt, mark size=3.7pt, mark=triangle, mark options={solid, rotate=180, mycolor1}]
  table[row sep=crcr]{%
0	0.757648095467463\\
3	0.651539179266971\\
6	0.533543755305448\\
9	0.415280259967464\\
12	0.307902176637543\\
15	0.2178913815621\\
18	0.146433634574631\\
21	0.092223535845004\\
24	0.053695877028911\\
27	0.0287704038090988\\
30	0.0142813549864342\\
33	0.00666043766857136\\
36	0.00296341639787228\\
39	0.00127351360888863\\
42	0.000533824491629385\\
45	0.000220525534993026\\
};
\addlegendentry{16-PSK}

\addplot [color=mycolor1, dash pattern={on 5pt off 3pt on 1pt off 3pt} , line width=0.7pt, mark size=3.7pt, mark=triangle, mark options={solid, rotate=90, mycolor1}]
  table[row sep=crcr]{%
0	0.571045038182275\\
3	0.49315935136821\\
6	0.4065040920897\\
9	0.318353875944136\\
12	0.235168232371592\\
15	0.161785517561926\\
18	0.102170078381952\\
21	0.0587760284993578\\
24	0.0309159066142407\\
27	0.0150647724709607\\
30	0.00691812960943586\\
33	0.00304163056083831\\
36	0.00129574930821978\\
39	0.000540400810333464\\
42	0.000222815062015642\\
45	9.11968165708369e-05\\
};
\addlegendentry{16-QAM}

\addplot [color=mycolor1, dash pattern={on 5pt off 3pt on 1pt off 3pt} , line width=0.7pt, mark size=3.7pt, mark=triangle, mark options={solid, rotate=270, mycolor1}]
  table[row sep=crcr]{%
0	0.895674327401571\\
3	0.778467964903144\\
6	0.64876762258788\\
9	0.518234943272379\\
12	0.397662840624723\\
15	0.293671756574982\\
18	0.20793781721924\\
21	0.139355490608538\\
24	0.0868431160119038\\
27	0.0497094446267155\\
30	0.0261290207214546\\
33	0.0127456487561774\\
36	0.00586139222759141\\
39	0.00258002080079685\\
42	0.00109985615747341\\
45	0.0004588020078266\\
};
\addlegendentry{64-QAM}

\addplot [color=red, draw=none, mark size=2.0pt, mark=asterisk, mark options={solid, red}]
  table[row sep=crcr]{%
0	0.475619623250311\\
3	0.448158110686024\\
6	0.397944542549645\\
9	0.319623410047207\\
12	0.223434979335542\\
15	0.134152689241089\\
18	0.0703366131498683\\
21	0.0332829024173976\\
24	0.0146965424410429\\
27	0.00619349902691856\\
30	0.00253374014717559\\
33	0.00102146313851809\\
36	0.000408570181131701\\
39	0.000160966476870366\\
42	6.08809739394978e-05\\
45	2.19660424578518e-05\\
};
\addlegendentry{Monte Carlo}

\addplot [color=red, draw=none, mark size=2.0pt, mark=asterisk, mark options={solid, red}, forget plot]
  table[row sep=crcr]{%
0	0.315768589748116\\
3	0.242069529758724\\
6	0.168328728428027\\
9	0.105299433137217\\
12	0.0594041684226517\\
15	0.0305962773614037\\
18	0.0146460039683937\\
21	0.00663568424441392\\
24	0.0028882316514401\\
27	0.00122180295185887\\
30	0.000507978570083546\\
33	0.000209403950144642\\
36	8.55360414775296e-05\\
39	3.43501664794626e-05\\
42	1.35251015939023e-05\\
45	5.3664090745105e-06\\
};
\addplot [color=red, draw=none, mark size=2.0pt, mark=asterisk, mark options={solid, red}, forget plot]
  table[row sep=crcr]{%
0	0.501708060665967\\
3	0.429071803615067\\
6	0.347534855460829\\
9	0.264324243123336\\
12	0.186962440895644\\
15	0.121655901075286\\
18	0.0722559602434827\\
21	0.03919415991085\\
24	0.0196083166109786\\
27	0.00918953636980934\\
30	0.00410114961759739\\
33	0.00176671856407097\\
36	0.000742397090255414\\
39	0.000307253191689341\\
42	0.000126016320745133\\
45	5.11876023008889e-05\\
};
\addplot [color=red, draw=none, mark size=2.0pt, mark=asterisk, mark options={solid, red}, forget plot]
  table[row sep=crcr]{%
0	0.757648095467463\\
3	0.651539179266971\\
6	0.533543755305448\\
9	0.415280259967464\\
12	0.307902176637543\\
15	0.2178913815621\\
18	0.146433634574631\\
21	0.092223535845004\\
24	0.053695877028911\\
27	0.0287704038090988\\
30	0.0142813549864342\\
33	0.00666043766857136\\
36	0.00296341639787228\\
39	0.00127351360888863\\
42	0.000533824491629385\\
45	0.000220525534993026\\
};
\addplot [color=red, draw=none, mark size=2.0pt, mark=asterisk, mark options={solid, red}, forget plot]
  table[row sep=crcr]{%
0	0.571045038182275\\
3	0.49315935136821\\
6	0.4065040920897\\
9	0.318353875944136\\
12	0.235168232371592\\
15	0.161785517561926\\
18	0.102170078381952\\
21	0.0587760284993578\\
24	0.0309159066142407\\
27	0.0150647724709607\\
30	0.00691812960943586\\
33	0.00304163056083831\\
36	0.00129574930821978\\
39	0.000540400810333464\\
42	0.000222815062015642\\
45	9.11968165708369e-05\\
};
\addplot [color=red, draw=none, mark size=2.0pt, mark=asterisk, mark options={solid, red}, forget plot]
  table[row sep=crcr]{%
0	0.895674327401571\\
3	0.778467964903144\\
6	0.64876762258788\\
9	0.518234943272379\\
12	0.397662840624723\\
15	0.293671756574982\\
18	0.20793781721924\\
21	0.139355490608538\\
24	0.0868431160119038\\
27	0.0497094446267155\\
30	0.0261290207214546\\
33	0.0127456487561774\\
36	0.00586139222759141\\
39	0.00258002080079685\\
42	0.00109985615747341\\
45	0.0004588020078266\\
};
\end{axis}
\end{tikzpicture}%
    \caption{Probability of bit error for SEL impairment and receive heterodyne detection technique. Simulation of different modulation schemes. }
\end{figure}
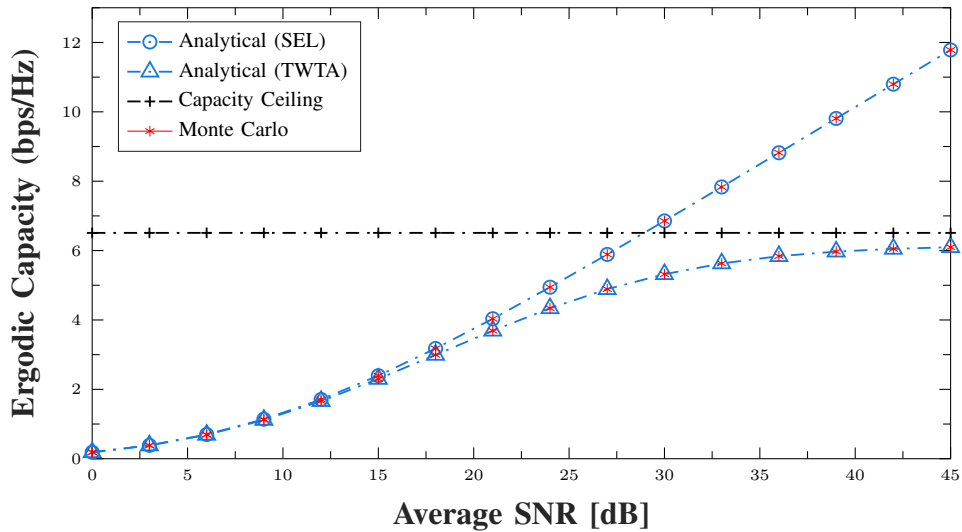
\begin{figure}[H]
\centering
\setlength\fheight{6cm}
\setlength\fwidth{12cm}
%
%
\definecolor{mycolor1}{rgb}{0.07059,0.48235,0.88627}%
\begin{tikzpicture}

\begin{axis}[%
width=0.951\fwidth,
height=\fheight,
at={(0\fwidth,0\fheight)},
scale only axis,
xmin=0,
xmax=45,
xlabel style={font=\bfseries\color{white!15!black}},
xlabel={Average SNR [dB]},
ymin=0,
ymax=13,
ylabel style={font=\bfseries\color{white!15!black}},
ylabel={Ergodic Capacity (bps/Hz)},
axis background/.style={fill=white},
legend style={at={(0.03,0.97)}, anchor=north west, legend cell align=left, align=left, draw=white!15!black}
]
\addplot [color=mycolor1, dash pattern={on 5pt off 3pt on 1pt off 3pt} , line width=0.7pt, mark size=2.5pt, mark=o, mark options={solid, mycolor1}]
  table[row sep=crcr]{%
0	0.186837331473901\\
3	0.385771934570917\\
6	0.698751281140982\\
9	1.13927081691044\\
12	1.70824070071998\\
15	2.39436202210066\\
18	3.17767692120313\\
21	4.03498834521461\\
24	4.94458600962351\\
27	5.88884769934164\\
30	6.85483739158011\\
33	7.8337006058395\\
36	8.81959373504621\\
39	9.80855926328656\\
42	10.7974551393233\\
45	11.7828377726832\\
};
\addlegendentry{Analytical (SEL)}

\addplot [color=mycolor1, dash pattern={on 5pt off 3pt on 1pt off 3pt} , line width=0.7pt, mark size=3.7pt, mark=triangle, mark options={solid, mycolor1}]
  table[row sep=crcr]{%
0	0.185415802857752\\
3	0.381792351948089\\
6	0.689081990372496\\
9	1.11765635608675\\
12	1.66229755394417\\
15	2.30043839132371\\
18	2.99326027130286\\
21	3.69007388809963\\
24	4.33669080382689\\
27	4.88786039732104\\
30	5.31904432188496\\
33	5.63007683975428\\
36	5.83890882137697\\
39	5.97096566249738\\
42	6.05052899366384\\
45	6.09662772638669\\
};
\addlegendentry{Analytical (TWTA)}

\addplot [color=black, dash pattern={on 5pt off 3pt on 1pt off 3pt} , line width=0.7pt, mark=+, mark options={solid, black}]
  table[row sep=crcr]{%
0	18.9364866925604\\
3	18.9364866925604\\
6	18.9364866925604\\
9	18.9364866925604\\
12	18.9364866925604\\
15	18.9364866925604\\
18	18.9364866925604\\
21	18.9364866925604\\
24	18.9364866925604\\
27	18.9364866925604\\
30	18.9364866925604\\
33	18.9364866925604\\
36	18.9364866925604\\
39	18.9364866925604\\
42	18.9364866925604\\
45	18.9364866925604\\
};
\addlegendentry{Capacity Ceiling}

\addplot [color=red, draw=none, mark size=2.0pt, mark=asterisk, mark options={solid, red}]
  table[row sep=crcr]{%
0	0.186837331473901\\
3	0.385771934570917\\
6	0.698751281140982\\
9	1.13927081691044\\
12	1.70824070071998\\
15	2.39436202210066\\
18	3.17767692120313\\
21	4.03498834521461\\
24	4.94458600962351\\
27	5.88884769934164\\
30	6.85483739158011\\
33	7.8337006058395\\
36	8.81959373504621\\
39	9.80855926328656\\
42	10.7974551393233\\
45	11.7828377726832\\
};
\addlegendentry{Monte Carlo}

\addplot [color=black, dash pattern={on 5pt off 3pt on 1pt off 3pt} , line width=0.7pt, mark=+, mark options={solid, black}, forget plot]
  table[row sep=crcr]{%
0	6.51043137112749\\
3	6.51043137112749\\
6	6.51043137112749\\
9	6.51043137112749\\
12	6.51043137112749\\
15	6.51043137112749\\
18	6.51043137112749\\
21	6.51043137112749\\
24	6.51043137112749\\
27	6.51043137112749\\
30	6.51043137112749\\
33	6.51043137112749\\
36	6.51043137112749\\
39	6.51043137112749\\
42	6.51043137112749\\
45	6.51043137112749\\
};
\addplot [color=red, draw=none, mark size=2.0pt, mark=asterisk, mark options={solid, red}, forget plot]
  table[row sep=crcr]{%
0	0.185415802857752\\
3	0.381792351948089\\
6	0.689081990372496\\
9	1.11765635608675\\
12	1.66229755394417\\
15	2.30043839132371\\
18	2.99326027130286\\
21	3.69007388809963\\
24	4.33669080382689\\
27	4.88786039732104\\
30	5.31904432188496\\
33	5.63007683975428\\
36	5.83890882137697\\
39	5.97096566249738\\
42	6.05052899366384\\
45	6.09662772638669\\
};
\end{axis}
\end{tikzpicture}%
    \caption{Comparison of the ergodic capacity under SEL and TWTA impairments. The relay employs fixed relaying gain while the receiver detects the signal following the heterodyne mode.}
\end{figure}
\vspace*{-1cm}
Fig.~8 presents the dependence of the ergodic capacity of FG relaying against the average SNR provided that the system suffers from either SEL or TWTA HPA impairment. For low average SNR, the system response to both SEL and TWTA is still acceptable as the impacts of the two HPA impairments are the same and also negligeable. Hence, in this SNR range, we can neglect the impacts of the SEL and TWTA and consider the system operating under linear relaying. However, as the average SNR increases and for a given IBO value equal to 10 dB, the impact of TWTA becomes more severe than SEL and this can be shown by the saturation of the capacity by an irreducible ceiling which is roughly 6.2 bps/Hz. Although, the relays's amplifiers for both SEL and TWTA are characterized by having the same IBO value, the system performance degrades substantially under the TWTA impairments. We also note that even for high SNR, the SEL impact is still acceptable on the system performance since the capacity is not limited by a ceiling or a floor at least in this SNR range (below 45 dB). It turned out that the system operates better in acceptable conditions under the SEL impairments than the TWTA for a given IBO level.
\vspace*{-0.5cm}
\begin{figure}[H]
\centering
\setlength\fheight{6cm}
\setlength\fwidth{12cm}
%
%
\definecolor{mycolor1}{rgb}{0.07059,0.48235,0.88627}%
\begin{tikzpicture}

\begin{axis}[%
width=0.951\fwidth,
height=\fheight,
at={(0\fwidth,0\fheight)},
scale only axis,
xmin=0,
xmax=45,
xlabel style={font=\bfseries\color{white!15!black}},
xlabel={Average SNR [dB]},
ymin=0,
ymax=13,
ylabel style={font=\bfseries\color{white!15!black}},
ylabel={Ergodic Capacity (bps/Hz)},
axis background/.style={fill=white},
legend style={at={(0.03,0.97)}, anchor=north west, legend cell align=left, align=left, draw=white!15!black}
]
\addplot [color=mycolor1, dash pattern={on 5pt off 3pt on 1pt off 3pt} , line width=0.7pt, mark size=2.5pt, mark=o, mark options={solid, mycolor1}]
  table[row sep=crcr]{%
0	0.20759497970146\\
3	0.446106999879726\\
6	0.829937680508719\\
9	1.35309425086979\\
12	1.96231408611983\\
15	2.57379271439519\\
18	3.10460599777153\\
21	3.50363523118943\\
24	3.76637837959519\\
27	3.92202390461146\\
30	4.00778492722556\\
33	4.05301248202121\\
36	4.07628846056629\\
39	4.08811307884096\\
42	4.09408017560425\\
45	4.09708114866214\\
};
\addlegendentry{Upper Bound (IBO = 0 dB)}

\addplot [color=mycolor1, dash pattern={on 5pt off 3pt on 1pt off 3pt} , line width=0.7pt, mark size=1.8pt, mark=square, mark options={solid, mycolor1}]
  table[row sep=crcr]{%
0	0.21378888581635\\
3	0.46384849642139\\
6	0.873068012057624\\
9	1.44658799383535\\
12	2.14822920390455\\
15	2.91449903347503\\
18	3.67295465378308\\
21	4.35297885130524\\
24	4.89815082771489\\
27	5.28480868778995\\
30	5.52900519723511\\
33	5.66973218146177\\
36	5.74600026391369\\
39	5.78585233343674\\
42	5.80626167100816\\
45	5.81660378565967\\
};
\addlegendentry{Upper Bound (IBO = 3 dB)}

\addplot [color=mycolor1,dash pattern={on 5pt off 3pt on 1pt off 3pt} , line width=0.7pt, mark size=4.3pt, mark=diamond, mark options={solid, mycolor1}]
  table[row sep=crcr]{%
0	0.215762174025031\\
3	0.46957490256395\\
6	0.887228380406904\\
9	1.47811599767841\\
12	2.21387835212416\\
15	3.04460361710863\\
18	3.91829004516251\\
21	4.78682732549383\\
24	5.60178769008896\\
27	6.31190320542373\\
30	6.87292209388936\\
33	7.26793165156168\\
36	7.51652001420708\\
39	7.65953310336097\\
42	7.73697591727446\\
45	7.77742539040851\\
};
\addlegendentry{Upper Bound (IBO = 5 dB)}

\addplot [color=mycolor1, dash pattern={on 5pt off 3pt on 1pt off 3pt} , line width=0.7pt, mark size=3.7pt, mark=triangle, mark options={solid, mycolor1}]
  table[row sep=crcr]{%
0	0.216420774227096\\
3	0.471494244510246\\
6	0.892001363899211\\
9	1.48884045262069\\
12	2.23658324522899\\
15	3.0910208675893\\
18	4.01093817237908\\
21	4.96753638581281\\
24	5.94360438065375\\
27	6.92966756429204\\
30	7.92068966806427\\
33	8.91395949328821\\
36	9.90784726251823\\
39	10.9010236305626\\
42	11.8918139746815\\
45	12.8773985166582\\
};
\addlegendentry{Upper Bound (IBO = 30 dB)}

\addplot [color=black, line width=1.0pt]
  table[row sep=crcr]{%
0	0.259857893636288\\
3	0.469744802587151\\
6	0.795407604158376\\
9	1.23796529276061\\
12	1.75762648102193\\
15	2.2893358574301\\
18	2.77360416301386\\
21	3.17513861481683\\
24	3.48363157076077\\
27	3.70593854077557\\
30	3.85747582401214\\
33	3.95582073239392\\
36	4.01692439692385\\
39	4.05346001691454\\
42	4.07458338646887\\
45	4.08643873525759\\
};
\addlegendentry{Approximation}

\addplot [color=black, dash pattern={on 5pt off 3pt on 1pt off 3pt} , line width=0.7pt, mark=+, mark options={solid, black}]
  table[row sep=crcr]{%
0	4.09959132248812\\
3	4.09959132248812\\
6	4.09959132248812\\
9	4.09959132248812\\
12	4.09959132248812\\
15	4.09959132248812\\
18	4.09959132248812\\
21	4.09959132248812\\
24	4.09959132248812\\
27	4.09959132248812\\
30	4.09959132248812\\
33	4.09959132248812\\
36	4.09959132248812\\
39	4.09959132248812\\
42	4.09959132248812\\
45	4.09959132248812\\
};
\addlegendentry{Capacity Ceiling}

\addplot [color=red, dash pattern={on 5pt off 3pt on 1pt off 3pt} , line width=0.7pt]
  table[row sep=crcr]{%
0	0.248130387548696\\
3	0.438128608376914\\
6	0.725253265431199\\
9	1.11262325995037\\
12	1.57543680476447\\
15	2.06537333437739\\
18	2.52776692151779\\
21	2.92047435332169\\
24	3.22393860682562\\
27	3.43987058367197\\
30	3.58307505529391\\
33	3.67261777071176\\
36	3.72595836871215\\
39	3.75650608214624\\
42	3.7734542255171\\
45	3.78262118043331\\
};
\addlegendentry{Exact}

\addplot [color=black, dash pattern={on 5pt off 3pt on 1pt off 3pt} , line width=0.7pt, mark=+, mark options={solid, black}, forget plot]
  table[row sep=crcr]{%
0	5.82653854398952\\
3	5.82653854398952\\
6	5.82653854398952\\
9	5.82653854398952\\
12	5.82653854398952\\
15	5.82653854398952\\
18	5.82653854398952\\
21	5.82653854398952\\
24	5.82653854398952\\
27	5.82653854398952\\
30	5.82653854398952\\
33	5.82653854398952\\
36	5.82653854398952\\
39	5.82653854398952\\
42	5.82653854398952\\
45	5.82653854398952\\
};
\addplot [color=red, dash pattern={on 5pt off 3pt on 1pt off 3pt} , line width=0.7pt, forget plot]
  table[row sep=crcr]{%
0	0.251735995275215\\
3	0.449331911155467\\
6	0.756238117486469\\
9	1.18781636641275\\
12	1.73488267960371\\
15	2.36169562181395\\
18	3.01414842982532\\
21	3.63370642851876\\
24	4.17213736520175\\
27	4.60204121197277\\
30	4.91922831854911\\
33	5.13717872184592\\
36	5.27793164552627\\
39	5.36417009681054\\
42	5.41474738081262\\
45	5.44336784955795\\
};
\addplot [color=black, dash pattern={on 5pt off 3pt on 1pt off 3pt} , line width=0.7pt, mark=+, mark options={solid, black}, forget plot]
  table[row sep=crcr]{%
0	7.81871113926323\\
3	7.81871113926323\\
6	7.81871113926323\\
9	7.81871113926323\\
12	7.81871113926323\\
15	7.81871113926323\\
18	7.81871113926323\\
21	7.81871113926323\\
24	7.81871113926323\\
27	7.81871113926323\\
30	7.81871113926323\\
33	7.81871113926323\\
36	7.81871113926323\\
39	7.81871113926323\\
42	7.81871113926323\\
45	7.81871113926323\\
};
\addplot [color=red, dash pattern={on 5pt off 3pt on 1pt off 3pt} , line width=0.7pt, forget plot]
  table[row sep=crcr]{%
0	0.252872866623861\\
3	0.452944330899478\\
6	0.766602081608767\\
9	1.21444750235516\\
12	1.7962927431069\\
15	2.48951259938411\\
18	3.25515468508374\\
21	4.04556159798313\\
24	4.81026726054521\\
27	5.50204526551228\\
30	6.08491037324724\\
33	6.54135565895626\\
36	6.87407837040222\\
39	7.10097688913169\\
42	7.24680563921773\\
45	7.33587823448313\\
};
\addplot [color=red, dash pattern={on 5pt off 3pt on 1pt off 3pt} , line width=0.7pt, forget plot]
  table[row sep=crcr]{%
0	0.253251081395348\\
3	0.454155089698488\\
6	0.770120784116535\\
9	1.2236910864144\\
12	1.81842182124519\\
15	2.53850258583727\\
18	3.35679911208576\\
21	4.24454851811788\\
24	5.17756337589957\\
27	6.13822295504842\\
30	7.11485147945996\\
33	8.10013930627597\\
36	9.0895341508833\\
39	10.079892216075\\
42	11.0683224663641\\
45	12.0509888733229\\
};
\addplot [color=black, line width=0.7pt, forget plot]
  table[row sep=crcr]{%
0	0.263926714537978\\
3	0.4831127630004\\
6	0.834223231799262\\
9	1.33442258802465\\
12	1.96045712282053\\
15	2.65364323858173\\
18	3.34413419344415\\
21	3.97304958250457\\
24	4.50353289033247\\
27	4.92194940087955\\
30	5.23283533103264\\
33	5.45165502137075\\
36	5.59823716647053\\
39	5.69209351998282\\
42	5.74978698439815\\
45	5.78398526401134\\
};
\addplot [color=black, line width=0.7pt, forget plot]
  table[row sep=crcr]{%
0	0.265212831632201\\
3	0.48745075397967\\
6	0.847388260917495\\
9	1.36947881866322\\
12	2.04188257218593\\
15	2.82009354102786\\
18	3.64688197433982\\
21	4.46681079034056\\
24	5.2313280348489\\
27	5.90242042476635\\
30	6.45728978874916\\
33	6.8902938865739\\
36	7.21020141536653\\
39	7.43471218558902\\
42	7.58489477136251\\
45	7.68100308414412\\
};
\addplot [color=black, line width=0.7pt, forget plot]
  table[row sep=crcr]{%
0	0.265641043278125\\
3	0.488907828418072\\
6	0.85188034890081\\
9	1.38177360821172\\
12	2.0718115433754\\
15	2.8861647249521\\
18	3.78193068085218\\
21	4.72515148369612\\
24	5.69412133959923\\
27	6.67644944457051\\
30	7.66540785590341\\
33	8.65730387636152\\
36	9.64984011353776\\
39	10.6410334018718\\
42	11.6282814421432\\
45	12.6072197595899\\
};
\end{axis}
\end{tikzpicture}%
    \caption{Ergodic capacity for different levels of IBO. The relay adopts the variable relaying gain under SEL impairment. The receiver detects the signal in heterodyne mode.}
\end{figure}
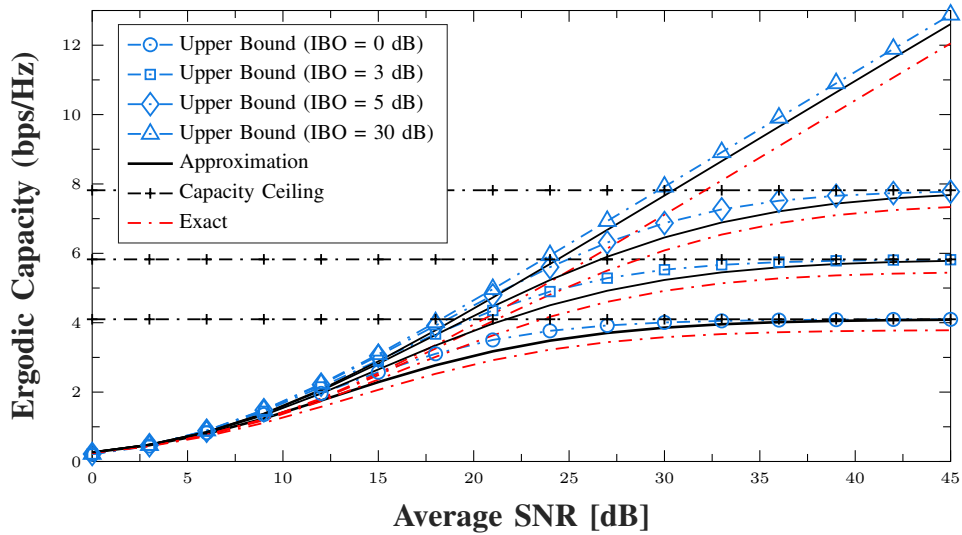
\begin{figure}[H]
\centering
\setlength\fheight{6cm}
\setlength\fwidth{12cm}
%
%
\definecolor{mycolor1}{rgb}{0.07059,0.48235,0.88627}%
\begin{tikzpicture}

\begin{axis}[%
width=0.951\fwidth,
height=\fheight,
at={(0\fwidth,0\fheight)},
scale only axis,
xmin=0,
xmax=50,
xlabel style={font=\bfseries\color{white!15!black}},
xlabel={Average SNR [dB]},
ymode=log,
ymin=1e-07,
ymax=1,
yminorticks=true,
ylabel style={font=\bfseries\color{white!15!black}},
ylabel={End-to-End Outage Probability},
axis background/.style={fill=white},
legend style={at={(0.03,0.03)}, anchor=south west, legend cell align=left, align=left, draw=white!15!black}
]
\addplot [color=mycolor1, dash pattern={on 5pt off 3pt on 1pt off 3pt} , line width=0.7pt, mark size=2.5pt, mark=o, mark options={solid, mycolor1}]
  table[row sep=crcr]{%
0	0.99946234992518\\
2.5	0.983196140847866\\
5	0.875815249674561\\
7.5	0.627425463738621\\
10	0.346784319857884\\
12.5	0.152029945474197\\
15	0.0556886128738812\\
17.5	0.0181897857638879\\
20	0.00578898215360779\\
22.5	0.0020307138787965\\
25	0.000893368701515129\\
27.5	0.000518282271734782\\
30	0.000375784375511889\\
32.5	0.000313478553643498\\
35	0.00028326884522889\\
37.5	0.000267650745413572\\
40	0.000259272576323522\\
42.5	0.000254683969612812\\
45	0.000252141522753657\\
47.5	0.000250723628771765\\
50	0.000249930000477017\\
};
\addlegendentry{SEL (IBO = 0 dB)}

\addplot [color=mycolor1, dash pattern={on 5pt off 3pt on 1pt off 3pt} , line width=0.7pt, mark size=3.7pt, mark=triangle, mark options={solid, mycolor1}]
  table[row sep=crcr]{%
0	0.999377631193507\\
2.5	0.981128948939443\\
5	0.864446769899423\\
7.5	0.603965503816914\\
10	0.321089395474843\\
12.5	0.133358820876676\\
15	0.0452609618031763\\
17.5	0.0131769352004734\\
20	0.00347981805810904\\
22.5	0.000895425687168605\\
25	0.00024548809130609\\
27.5	7.81201791585362e-05\\
30	3.00306298758324e-05\\
32.5	1.37078898160281e-05\\
35	7.15254139338484e-06\\
37.5	4.15484562810153e-06\\
40	2.66199678800039e-06\\
42.5	1.87917285232597e-06\\
45	1.45612655766403e-06\\
47.5	1.22352793485714e-06\\
50	1.09438040929444e-06\\
};
\addlegendentry{SEL (IBO = 4 dB)}

\addplot [color=mycolor1, dash pattern={on 5pt off 3pt on 1pt off 3pt} , line width=0.7pt, mark size=1.8pt, mark=square, mark options={solid, mycolor1}]
  table[row sep=crcr]{%
0	0.999791688895845\\
2.5	0.992177234284698\\
5	0.931811700600507\\
7.5	0.761538794546428\\
10	0.522366391736437\\
12.5	0.311123633390873\\
15	0.173553811217618\\
17.5	0.0995464139943562\\
20	0.0635948820420261\\
22.5	0.0466545135216552\\
25	0.0385219687462254\\
27.5	0.0344570278188852\\
30	0.0323420250601749\\
32.5	0.0312078567759312\\
35	0.0305875885191979\\
37.5	0.0302443179866267\\
40	0.0300530276541774\\
42.5	0.0299460078196317\\
45	0.0298860000212764\\
47.5	0.0298523100931952\\
50	0.029833382217917\\
};
\addlegendentry{SEL (IBO = 6 dB)}

\addplot [color=mycolor1, dash pattern={on 5pt off 3pt on 1pt off 3pt} , line width=0.7pt, mark size=4.3pt, mark=diamond, mark options={solid, mycolor1}]
  table[row sep=crcr]{%
0	0.999555121080559\\
2.5	0.985552090033769\\
5	0.889364878166195\\
7.5	0.656881280209739\\
10	0.381079928053685\\
12.5	0.178885636043667\\
15	0.0722118468864541\\
17.5	0.0272379143420174\\
20	0.0107377625134429\\
22.5	0.00500917237377752\\
25	0.00296162913185938\\
27.5	0.00215794215326137\\
30	0.00180437198457029\\
32.5	0.00163333379019615\\
35	0.00154517585179592\\
37.5	0.00149798245179456\\
40	0.00147216697194275\\
42.5	0.00145787311867385\\
45	0.0014499047058576\\
47.5	0.00144544557415505\\
50	0.00144294489377383\\
};
\addlegendentry{SEL (IBO = 8 dB)}

\addplot [color=black, dash pattern={on 5pt off 3pt on 1pt off 3pt} , line width=0.7pt, mark=+, mark options={solid, black}]
  table[row sep=crcr]{%
0	0.000249930000477017\\
2.5	0.000249930000477017\\
5	0.000249930000477017\\
7.5	0.000249930000477017\\
10	0.000249930000477017\\
12.5	0.000249930000477017\\
15	0.000249930000477017\\
17.5	0.000249930000477017\\
20	0.000249930000477017\\
22.5	0.000249930000477017\\
25	0.000249930000477017\\
27.5	0.000249930000477017\\
30	0.000249930000477017\\
32.5	0.000249930000477017\\
35	0.000249930000477017\\
37.5	0.000249930000477017\\
40	0.000249930000477017\\
42.5	0.000249930000477017\\
45	0.000249930000477017\\
47.5	0.000249930000477017\\
50	0.000249930000477017\\
};
\addlegendentry{Numerical Outage Floor}

\addplot [color=red, draw=none, mark size=2.0pt, mark=asterisk, mark options={solid, red}]
  table[row sep=crcr]{%
0	0.99946234992518\\
2.5	0.983196140847866\\
5	0.875815249674561\\
7.5	0.627425463738621\\
10	0.346784319857884\\
12.5	0.152029945474197\\
15	0.0556886128738812\\
17.5	0.0181897857638879\\
20	0.00578898215360779\\
22.5	0.0020307138787965\\
25	0.000893368701515129\\
27.5	0.000518282271734782\\
30	0.000375784375511889\\
32.5	0.000313478553643498\\
35	0.00028326884522889\\
37.5	0.000267650745413572\\
40	0.000259272576323522\\
42.5	0.000254683969612812\\
45	0.000252141522753657\\
47.5	0.000250723628771765\\
50	0.000249930000477017\\
};
\addlegendentry{Monte Carlo}

\addplot [color=black, dash pattern={on 5pt off 3pt on 1pt off 3pt} , line width=0.7pt, mark=+, mark options={solid, black}, forget plot]
  table[row sep=crcr]{%
0	1.09438040929444e-06\\
2.5	1.09438040929444e-06\\
5	1.09438040929444e-06\\
7.5	1.09438040929444e-06\\
10	1.09438040929444e-06\\
12.5	1.09438040929444e-06\\
15	1.09438040929444e-06\\
17.5	1.09438040929444e-06\\
20	1.09438040929444e-06\\
22.5	1.09438040929444e-06\\
25	1.09438040929444e-06\\
27.5	1.09438040929444e-06\\
30	1.09438040929444e-06\\
32.5	1.09438040929444e-06\\
35	1.09438040929444e-06\\
37.5	1.09438040929444e-06\\
40	1.09438040929444e-06\\
42.5	1.09438040929444e-06\\
45	1.09438040929444e-06\\
47.5	1.09438040929444e-06\\
50	1.09438040929444e-06\\
};
\addplot [color=black, dash pattern={on 5pt off 3pt on 1pt off 3pt} , line width=0.7pt, mark=+, mark options={solid, black}, forget plot]
  table[row sep=crcr]{%
0	0.029833382217917\\
2.5	0.029833382217917\\
5	0.029833382217917\\
7.5	0.029833382217917\\
10	0.029833382217917\\
12.5	0.029833382217917\\
15	0.029833382217917\\
17.5	0.029833382217917\\
20	0.029833382217917\\
22.5	0.029833382217917\\
25	0.029833382217917\\
27.5	0.029833382217917\\
30	0.029833382217917\\
32.5	0.029833382217917\\
35	0.029833382217917\\
37.5	0.029833382217917\\
40	0.029833382217917\\
42.5	0.029833382217917\\
45	0.029833382217917\\
47.5	0.029833382217917\\
50	0.029833382217917\\
};
\addplot [color=black, dash pattern={on 5pt off 3pt on 1pt off 3pt} , line width=0.7pt, mark=+, mark options={solid, black}, forget plot]
  table[row sep=crcr]{%
0	0.00144294489377383\\
2.5	0.00144294489377383\\
5	0.00144294489377383\\
7.5	0.00144294489377383\\
10	0.00144294489377383\\
12.5	0.00144294489377383\\
15	0.00144294489377383\\
17.5	0.00144294489377383\\
20	0.00144294489377383\\
22.5	0.00144294489377383\\
25	0.00144294489377383\\
27.5	0.00144294489377383\\
30	0.00144294489377383\\
32.5	0.00144294489377383\\
35	0.00144294489377383\\
37.5	0.00144294489377383\\
40	0.00144294489377383\\
42.5	0.00144294489377383\\
45	0.00144294489377383\\
47.5	0.00144294489377383\\
50	0.00144294489377383\\
};
\addplot [color=red, draw=none, mark size=2.0pt, mark=asterisk, mark options={solid, red}, forget plot]
  table[row sep=crcr]{%
0	0.999377631193507\\
2.5	0.981128948939443\\
5	0.864446769899423\\
7.5	0.603965503816914\\
10	0.321089395474843\\
12.5	0.133358820876676\\
15	0.0452609618031763\\
17.5	0.0131769352004734\\
20	0.00347981805810904\\
22.5	0.000895425687168605\\
25	0.00024548809130609\\
27.5	7.81201791585362e-05\\
30	3.00306298758324e-05\\
32.5	1.37078898160281e-05\\
35	7.15254139338484e-06\\
37.5	4.15484562810153e-06\\
40	2.66199678800039e-06\\
42.5	1.87917285232597e-06\\
45	1.45612655766403e-06\\
47.5	1.22352793485714e-06\\
50	1.09438040929444e-06\\
};
\addplot [color=red, draw=none, mark size=2.0pt, mark=asterisk, mark options={solid, red}, forget plot]
  table[row sep=crcr]{%
0	0.999791688895845\\
2.5	0.992177234284698\\
5	0.931811700600507\\
7.5	0.761538794546428\\
10	0.522366391736437\\
12.5	0.311123633390873\\
15	0.173553811217618\\
17.5	0.0995464139943562\\
20	0.0635948820420261\\
22.5	0.0466545135216552\\
25	0.0385219687462254\\
27.5	0.0344570278188852\\
30	0.0323420250601749\\
32.5	0.0312078567759312\\
35	0.0305875885191979\\
37.5	0.0302443179866267\\
40	0.0300530276541774\\
42.5	0.0299460078196317\\
45	0.0298860000212764\\
47.5	0.0298523100931952\\
50	0.029833382217917\\
};
\addplot [color=red, draw=none, mark size=2.0pt, mark=asterisk, mark options={solid, red}, forget plot]
  table[row sep=crcr]{%
0	0.999555121080559\\
2.5	0.985552090033769\\
5	0.889364878166195\\
7.5	0.656881280209739\\
10	0.381079928053685\\
12.5	0.178885636043667\\
15	0.0722118468864541\\
17.5	0.0272379143420174\\
20	0.0107377625134429\\
22.5	0.00500917237377752\\
25	0.00296162913185938\\
27.5	0.00215794215326137\\
30	0.00180437198457029\\
32.5	0.00163333379019615\\
35	0.00154517585179592\\
37.5	0.00149798245179456\\
40	0.00147216697194275\\
42.5	0.00145787311867385\\
45	0.0014499047058576\\
47.5	0.00144544557415505\\
50	0.00144294489377383\\
};
\end{axis}
\end{tikzpicture}%
    \caption{Probability of outage for different levels of IBO. The relay employs variable relaying gain under SEL impairment while IM/DD detection is assumed. The numerical outage floor is considered in this simulation for different IBO thresholds.}
\end{figure}
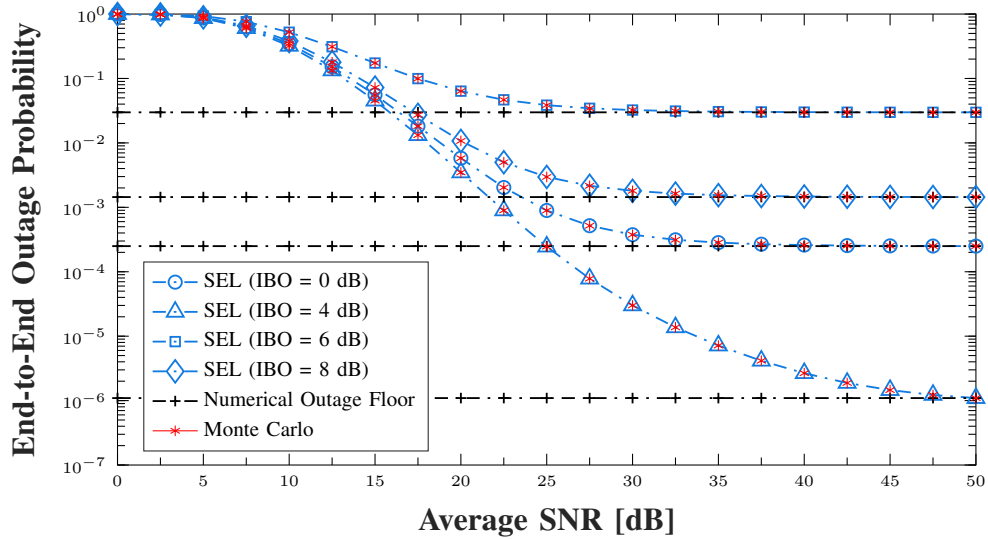
\vspace*{-1cm}
Fig.~9 shows the variations of the ergodic capacity of VG relaying versus the average SNR for different values of IBO. The relays suffer from the SEL impairments and the receiver detects the signal following IM/DD. Clearly, the three characteristics of the ergodic capacity, which are the exact Monte Carlo, approximate and upper bound, deviate from each other for low SNR but they overlap asymptotically at high SNR. Although the approximation given by (44) has no theoretical foundations, it is more tighter to the exact capacity compared to the upper bound derived from the Jensen's inequality. Graphically, we observe that the ergodic capacity saturates by the hardware ceilings created by the HPA non-linearities as shown by Fig.~(9). In addition, these ceilings disappear for an IBO = 30 dB but the performance is limited for the  case of lower values of IBO. For the following values of IBO equal to 0, 3 and 5 dB, the system capacity is saturated by the following ceiling values 4, 5.9 and 7.9 bps/Hz, respectively. Note that these ceilings are inversely proportional to the values of the IBO. In fact, as the IBO increases, the saturation amplitude of the relay amplifier increases and thereby the distortion effect is reduced. However, as the IBO decreases, i.e, the relay amplifier level becomes lower, the nonlinear distortion impact becomes more severe and the channel capacity substantially saturates. Note that the capacity ceiling depends only on the hardware impairment parameters like the clipping factor and the scale of the input signal and not on the system parameters as the number of the relays and the channel parameters. Hence, it is straightforward that for any system suffering from the hardware impairments, the channel capacity is always limited by the impairment ceiling regardless of the system configuration such as the channels nature (RF/FSO) and the number of the relays, etc.\\
Fig.~10 illustrates the variations of the outage probability with respect to the IBO. Clearly, we observe the distortion impact on the performance by the creation of the outage floor. This floor essentially saturates the system since the relay amplifier is not able to provide the required power and as a result a clipping and a distortion both affect the signal. As the IBO increases, i.e, the saturation level of the relay enhances and the amplifier can provide higher amount of power. Consequently, the impacts of the distortion and the clipping on the signal are mitigated. This case is importantly manifested for IBO = 8 dB, although, the inevitable outage floor is created, the system is still working better than for the case of lower IBO values.
\vspace*{-0.5cm}
\section{Conclusion}
In this paper, we investigate a mixed RF/FSO systems with multiple relays where the RF and FSO links are modeled by Rayleigh and unified Double Generalized Gamma distributions, respectively. PRS with outdated CSI is assumed to select one relay among the set since the channels are time-varying and hence the selection is primarily based on the outdated CSI due to the propagation delay. We also consider FG and VG relaying schemes for the global performance analysis and we introduce the SEL and TWTA HPA non-linearities to the relays that occur during the amplification. We derive new closed-forms of the OP, BEP and ergodic capacity and we also evaluate the asymptotic performance of the system at high SNR regime. We show that the system works better for weak optical fading such as the turbulence-induced fading, the atmospheric path loss and the pointing error fading, howerver, as the FSO fading becomes severe, the performance gets worse even for better system configuration. We also prove that a better correlation of the CSIs yields a lower outage performance. Additionally, the analysis of the ergodic capacity shows that for a given IBO, the effect of the HPA non-linearities can be neglected and the system can be considered operating under linear relaying regime. However, as the average SNR increases, the system performance becomes very sensitive to the TWTA and the capacity saturates quickly around 30 dB than for the SEL impairments. Practically, the SEL amplifier is shown to be more efficient than TWT amplifier since it allows the system to operate in acceptable condition for the same amount of IBO. Furthermore, further investigation of the ergodic capacity for CSI-assisted relaying prove that the system rate substantially improves as the IBO increases since the constraint on the peak power during the amplification allow the amplifier to provide higher power to the signal without clipping the signal peaks. 
\vspace*{-0.5cm}
\appendices
\section{Derivation of the Outage Probability for Fixed Gain Relaying}
After replacing (27) in (29), the outage probability can be written as
\begin{equation}
\begin{split}
P_{\text{out}}(\gamma_{\text{th}}) = \text{Pr}\left[\frac{\gamma_{1(m)}\gamma_{2(m)}}{\kappa\gamma_{2(m)} + c} < \gamma_{\text{th}}\right] = \text{Pr}\left[\gamma_{1(m)} < \gamma_{\text{th}}\kappa + \frac{c\gamma_{\text{th}}}{\gamma_{2(m)}}   \right],
\end{split}    
\end{equation}
Since the RF and FSO fadings are independent, the OP can be expressed as follows
\begin{equation}
\begin{split}
P_{\text{out}}(\gamma_{\text{th}}) = \int\limits_0^\infty F_{\gamma_{1(m)}}\left(\kappa\gamma_{\text{th}} + \frac{c\gamma_{\text{th}}}{\gamma} \right)f_{\gamma_{2(m)}}(\gamma)~d\gamma,
\end{split}    
\end{equation}
After changing the variable of the integration ($x = \gamma^{-1}$), we transform the exponential function to Meijer's-G function \cite[Eq.~(07.34.03.0046.01)]{68}. After using the identity \cite[Eq.~(2.24.1.1)]{62}, the OP can be derived as (30).
\vspace*{-1cm}
\section{Derivation of the Capacity Upper Bound for Variable Gain Relaying}
Since the derivation of the term $\mathcal{J}$ using the expression of the end-to-end SNDR (43) is not tractable, we consider an approximation of (43). In this context, we define the term $\mathcal{J}$ as follows
\begin{equation}
\begin{split}
\mathcal{J}  &\cong \e{\frac{\gamma_{1(m)}\gamma_{2(m)}}{\kappa\gamma_{2(m)} +  \gamma_{1(m)}}}
= \int\limits_0^\infty \gamma_{2}f_{\gamma_{2(m)}}(\gamma_2)\underbrace{\left[\int\limits_0^\infty\frac{\gamma_1}{\kappa\gamma_2+\gamma_1}f_{\gamma_{1(m)}}(\gamma_1)~d\gamma_1\right]}_{\mathcal{I}}d\gamma_2,
\end{split}
\end{equation}
Using the integral identity \cite[Eq.~(2.3.6.13)]{67}, the term $\mathcal{I}$ can be derived in term of the incomplete upper Gamma function. The next step is to transform the exponential, the incomplete upper Gamma function and the Meijer's-G into Fox-H function as follows
\begin{equation}
\begin{split}
G_{p,q}^{m,n}\Bigg[z^C~\bigg|~\begin{matrix} a_1,\ldots,a_p \\ b_1,\ldots,b_q \end{matrix} \Bigg] = \frac{1}{C}~H_{p,q}^{m,n}\Bigg[z~\bigg|~\begin{matrix} (a_1,C^{-1}),\ldots,(a_p,C^{-1}) \\ (b_1,C^{-1}),\ldots,(b_q,C^{-1}) \end{matrix} \Bigg],
\end{split}
\end{equation}
Then, we transform both the exponential and the upper incomplete Gamma function into Meijer's-G functions\cite[Eq.~(06.06.26.0005.01)]{68} and then we transform them into Fox-H function as follows
\vspace*{-0.75cm}
\begin{equation}
\begin{split}
&e^{-a\gamma} =  G_{0,1}^{1,0} \Bigg[a\gamma~\bigg|~\begin{matrix} - \\ 0 \end{matrix} \Bigg] = H_{0,1}^{1,0} \Bigg[a\gamma~\bigg|~\begin{matrix} - \\ (0,1) \end{matrix} \Bigg],  
\end{split}    
\end{equation}
\vspace*{-0.5cm}
\begin{equation}
\begin{split}
&\Gamma(\alpha,x) =  G_{1,2}^{2,0} \Bigg[x~\bigg|~\begin{matrix} 1 \\ \alpha, 0 \end{matrix} \Bigg] = H_{1,2}^{2,0} \Bigg[x~\bigg|~\begin{matrix} (1,1) \\ (\alpha,1), (0,1) \end{matrix} \Bigg],  
\end{split}    
\end{equation}
\vspace*{-0.25cm}
After transforming the Meijer's-G function involved in the expression of the Double Generalized Gamma fading into Fox-H function using (67), the integral involves three Fox-H functions. After applying the identity \cite[Eq.~(2.3)]{63}, the term $\mathcal{J}$ is finally derived.
\vspace*{-0.5cm}
\ifCLASSOPTIONcaptionsoff
  \newpage
\fi
\bibliographystyle{IEEEtran}
\bibliography{main}

\begin{thebibliography}{10}
\providecommand{\url}[1]{#1}
\csname url@samestyle\endcsname
\providecommand{\newblock}{\relax}
\providecommand{\bibinfo}[2]{#2}
\providecommand{\BIBentrySTDinterwordspacing}{\spaceskip=0pt\relax}
\providecommand{\BIBentryALTinterwordstretchfactor}{4}
\providecommand{\BIBentryALTinterwordspacing}{\spaceskip=\fontdimen2\font plus
\BIBentryALTinterwordstretchfactor\fontdimen3\font minus
  \fontdimen4\font\relax}
\providecommand{\BIBforeignlanguage}[2]{{%
\expandafter\ifx\csname l@#1\endcsname\relax
\typeout{** WARNING: IEEEtran.bst: No hyphenation pattern has been}%
\typeout{** loaded for the language `#1'. Using the pattern for}%
\typeout{** the default language instead.}%
\else
\language=\csname l@#1\endcsname
\fi
#2}}
\providecommand{\BIBdecl}{\relax}
\BIBdecl

\bibitem{5}
M.~A. Khalighi and M.~Uysal, ``{S}urvey on {F}ree {S}pace {O}ptical
  {C}ommunication: {A} {C}ommunication {T}heory {P}erspective,'' \emph{IEEE
  Communications Surveys Tutorials}, vol.~16, no.~4, pp. 2231--2258,
  Fourthquarter 2014.

\bibitem{8}
X.~Tang, Z.~Wang, Z.~Xu, and Z.~Ghassemlooy, ``{M}ultihop {F}ree-{S}pace
  {O}ptical {C}ommunications {O}ver {T}urbulence {C}hannels with {P}ointing
  {E}rrors using {H}eterodyne {D}etection,'' \emph{Journal of Lightwave
  Technology}, vol.~32, no.~15, pp. 2597--2604, Aug 2014.

\bibitem{10}
X.~Ge, S.~Tu, G.~Mao, C.~X. Wang, and T.~Han, ``{5G} {U}ltra-{D}ense {C}ellular
  {N}etworks,'' \emph{IEEE Wireless Communications}, vol.~23, no.~1, pp.
  72--79, February 2016.

\bibitem{mmwave}
P.~V. Trinh, T.~C. Thang, and A.~T. Pham, ``Mixed mmwave {{RF/FSO}} relaying
  systems over generalized fading channels with pointing errors,'' \emph{IEEE
  Photonics Journal}, vol.~9, no.~1, pp. 1--14, Feb 2017.

\bibitem{ln1}
A.~A. Farid and S.~Hranilovic, ``Outage capacity optimization for free-space
  optical links with pointing errors,'' \emph{J. Lightwave Technol.}, vol.~25,
  no.~7, pp. 1702--1710, Jul 2007.

\bibitem{60}
E.~Balti, M.~Guizani, B.~Hamdaoui, and Y.~Maalej, ``Partial relay selection for
  hybrid {{RF/FSO}} systems with hardware impairments,'' in \emph{2016 IEEE
  Global Communications Conference (GLOBECOM)}, Dec 2016, pp. 1--6.

\bibitem{dgg}
M.~A. Kashani, M.~Uysal, and M.~Kavehrad, ``A novel statistical channel model
  for turbulence-induced fading in free-space optical systems,'' \emph{Journal
  of Lightwave Technology}, vol.~33, no.~11, pp. 2303--2312, June 2015.

\bibitem{43}
A.~Mansour, R.~Mesleh, and M.~Abaza, ``New challenges in wireless and free
  space optical communications,'' \emph{Optics and Lasers in Engineering},
  vol.~89, pp. 95 -- 108, 2017, 3DIM-DS 2015: Optical Image Processing in the
  context of 3D Imaging, Metrology, and Data Security.

\bibitem{44}
{Murat Uysal, Carlo Capsoni, Zabih Ghassemlooy, Anthony Boucouvalas, Eszter
  Udvary}, \emph{Optical Wireless Communications: An Emerging Technology},
  1st~ed., ser. Signals and Communication Technology.\hskip 1em plus 0.5em
  minus 0.4em\relax Springer International Publishing, 2016.

\bibitem{45}
F.~Yang, J.~Cheng, and T.~A. Tsiftsis, ``Free-space optical communication with
  nonzero boresight pointing errors,'' \emph{IEEE Transactions on
  Communications}, vol.~62, no.~2, pp. 713--725, February 2014.

\bibitem{46}
W.~Gappmair, S.~Hranilovic, and E.~Leitgeb, ``{OOK} performance for terrestrial
  {FSO} links in turbulent atmosphere with pointing errors modeled by {H}oyt
  distributions,'' \emph{IEEE Communications Letters}, vol.~15, no.~8, pp.
  875--877, August 2011.

\bibitem{47}
A.~A. Farid and S.~Hranilovic, ``Diversity gain and outage probability for mimo
  free-space optical links with misalignment,'' \emph{IEEE Transactions on
  Communications}, vol.~60, no.~2, pp. 479--487, February 2012.

\bibitem{48}
E.~Soleimani-Nasab and M.~Uysal, ``Generalized performance analysis of mixed
  {{RF/FSO}} cooperative systems,'' \emph{IEEE Transactions on Wireless
  Communications}, vol.~15, no.~1, pp. 714--727, Jan 2016.

\bibitem{int}
E.~Balti and M.~Guizani, ``Mixed rf/fso cooperative relaying systems with
  co-channel interference,'' \emph{IEEE Transactions on Communications}, pp.
  1--1, 2018.

\bibitem{2}
Z.~Ghassemlooy, W.~Popoola, and S.~Rajbhandari, \emph{{O}ptical {W}ireless
  {C}ommunications: {S}ystem and {C}hannel {M}odelling with {MATLAB}},
  1st~ed.\hskip 1em plus 0.5em minus 0.4em\relax Boca Raton, FL, USA: CRC
  Press, Inc., 2012.

\bibitem{50}
E.~Zedini, H.~Soury, and M.~S. Alouini, ``On the performance analysis of
  dual-hop mixed {FSO/RF} systems,'' \emph{IEEE Transactions on Wireless
  Communications}, vol.~15, no.~5, pp. 3679--3689, May 2016.

\bibitem{11}
C.~Hoymann, W.~Chen, J.~Montojo, A.~Golitschek, C.~Koutsimanis, and X.~Shen,
  ``{R}elaying {o}peration in {3GPP LTE}: {C}hallenges and {S}olutions,''
  \emph{IEEE Communications Magazine}, vol.~50, no.~2, pp. 156--162, February
  2012.

\bibitem{aggregate}
E.~Balti, M.~Guizani, B.~Hamdaoui, and B.~Khalfi, ``Aggregate hardware
  impairments over mixed rf/fso relaying systems with outdated csi,''
  \emph{IEEE Transactions on Communications}, vol.~PP, no.~99, pp. 1--1, 2017.

\bibitem{glob2017}
------, ``Mixed rf/fso relaying systems with hardware impairments,'' in
  \emph{GLOBECOM 2017 - 2017 IEEE Global Communications Conference}, Dec 2017,
  pp. 1--6.

\bibitem{17}
H.~Samimi and M.~Uysal, ``End-to-end performance of mixed {{RF/FSO}}
  transmission systems,'' \emph{IEEE/OSA Journal of Optical Communications and
  Networking}, vol.~5, no.~11, pp. 1139--1144, Nov 2013.

\bibitem{19}
S.~Anees and M.~R. Bhatnagar, ``Performance evaluation of decode-and-forward
  dual-hop asymmetric radio frequency-free space optical communication
  system,'' \emph{IET Optoelectronics}, vol.~9, no.~5, pp. 232--240, 2015.

\bibitem{20}
K.~Kumar and D.~K. Borah, ``Quantize and encode relaying through {FSO} and
  hybrid {FSO/RF} links,'' \emph{IEEE Transactions on Vehicular Technology},
  vol.~64, no.~6, pp. 2361--2374, June 2015.

\bibitem{21}
I.~Avram, N.~Aerts, H.~Bruneel, and M.~Moeneclaey, ``Quantize and forward
  cooperative communication: Channel parameter estimation,'' \emph{IEEE
  Transactions on Wireless Communications}, vol.~11, no.~3, pp. 1167--1179,
  March 2012.

\bibitem{24}
N.~Sharma, A.~Bansal, and P.~Garg, ``Relay selection in mixed {RF/FSO} system
  over generalized channel fading,'' \emph{Transactions on Emerging
  Telecommunications Technologies}, 2016.

\bibitem{27}
E.~Balti and M.~Guizani, ``{Impact of Non-Linear High Power Amplifiers on
  Cooperative Relaying Systems},'' \emph{IEEE Transactions on Communications},
  vol.~PP, no.~99, pp. 1--1, 2017.

\bibitem{28}
N.~Maletic, M.~Cabarkapa, and N.~Neskovic, ``Performance of fixed-gain
  amplify-and-forward nonlinear relaying with hardware impairments,''
  \emph{International Journal of Communication Systems}, 2015.

\bibitem{30}
J.~Li, M.~Matthaiou, and T.~Svensson, ``{I/Q} imbalance in {AF} dual-hop
  relaying: Performance analysis in {N}akagami-m fading,'' \emph{IEEE
  Transactions on Communications}, vol.~62, no.~3, pp. 836--847, March 2014.

\bibitem{31}
J.~Qi, S.~Aissa, and M.~S. Alouini, ``Analysis and compensation of {I/Q}
  imbalance in amplify-and-forward cooperative systems,'' in \emph{2012 IEEE
  Wireless Communications and Networking Conference (WCNC)}, April 2012, pp.
  215--220.

\bibitem{32}
E.~Bjornson, M.~Matthaiou, and M.~Debbah, ``A new look at dual-hop relaying:
  Performance limits with hardware impairments,'' \emph{IEEE Transactions on
  Communications}, vol.~61, no.~11, pp. 4512--4525, November 2013.

\bibitem{33}
M.~Matthaiou, A.~Papadogiannis, E.~Bjornson, and M.~Debbah, ``Two-way relaying
  under the presence of relay transceiver hardware impairments,'' \emph{IEEE
  Communications Letters}, vol.~17, no.~6, pp. 1136--1139, June 2013.

\bibitem{42}
J.~Li and J.~Ilow, ``Adaptive volterra predistorters for compensation of
  non-linear effects with memory in ofdm transmitters,'' in \emph{4th Annual
  Communication Networks and Services Research Conference (CNSR'06)}, May 2006,
  pp. 4 pp.--103.

\bibitem{41}
N.~Maletic, M.~Cabarkapa, and N.~Neskovic, ``Performance of fixed-gain
  amplify-and-forward nonlinear relaying with hardware impairments,''
  \emph{International Journal of Communication Systems}, 2015.

\bibitem{55}
E.~Zedini, I.~S. Ansari, and M.~S. Alouini, ``Performance analysis of mixed
  {N}akagami-m and {G}amma-{G}amma dual-hop {FSO} transmission systems,''
  \emph{IEEE Photonics Journal}, vol.~7, no.~1, pp. 1--20, Feb 2015.

\bibitem{56}
H.~AlQuwaiee, I.~S. Ansari, and M.~S. Alouini, ``On the performance of
  free-space optical communication systems over double generalized gamma
  channel,'' \emph{IEEE Journal on Selected Areas in Communications}, vol.~33,
  no.~9, pp. 1829--1840, Sept 2015.

\bibitem{58}
I.~S. Ansari, F.~Yilmaz, and M.~S. Alouini, ``On the performance of hybrid {RF}
  and {{RF/FSO}} dual-hop transmission systems,'' in \emph{2013 2nd
  International Workshop on Optical Wireless Communications (IWOW)}, Oct 2013,
  pp. 45--49.

\bibitem{59}
L.~Yang, M.~O. Hasna, and X.~Gao, ``Performance of mixed {{RF/FSO}} with
  variable gain over generalized atmospheric turbulence channels,'' \emph{IEEE
  Journal on Selected Areas in Communications}, vol.~33, no.~9, pp. 1913--1924,
  Sept 2015.

\bibitem{61}
E.~Balti, M.~Guizani, and B.~Hamdaoui, ``Hybrid rayleigh and {Double-Weibull}
  over impaired {{RF/FSO}} system with outdated {CSI},'' in \emph{IEEE ICC 2017
  Mobile and Wireless Networking (ICC'17 MWN)}, Paris, France, May 2017, pp.
  2093--2098.

\bibitem{jakes}
W.~C. Jakes and D.~C. Cox, Eds., \emph{Microwave Mobile Communications}.\hskip
  1em plus 0.5em minus 0.4em\relax Wiley-IEEE Press, 1994.

\bibitem{ofdm}
H.~Bouhadda, H.~Shaiek, D.~Roviras, R.~Zayani, Y.~Medjahdi, and R.~Bouallegue,
  ``Theoretical analysis of {BER} performance of nonlinearly amplified
  {FBMC/OQAM} and {OFDM} signals,'' \emph{EURASIP Journal on Advances in Signal
  Processing}, vol. 2014, no.~1, p.~60, 2014.

\bibitem{66}
M.~I. Petkovic, A.~M. Cvetkovic, G.~T. Djordjevic, and G.~K. Karagiannidis,
  ``Partial relay selection with outdated channel state estimation in mixed
  {{RF/FSO}} systems,'' \emph{Journal of Lightwave Technology}, vol.~33,
  no.~13, pp. 2860--2867, July 2015.

\bibitem{64}
I.~S. Gradshteyn and I.~M. Ryzhik, \emph{Table of integrals, series, and
  products}, 7th~ed.\hskip 1em plus 0.5em minus 0.4em\relax Elsevier/Academic
  Press, Amsterdam, 2007.

\bibitem{49}
G.~T. Djordjevic, M.~I. Petkovic, A.~M. Cvetkovic, and G.~K. Karagiannidis,
  ``Mixed {{RF/FSO}} relaying with outdated channel state information,''
  \emph{IEEE Journal on Selected Areas in Communications}, vol.~33, no.~9, pp.
  1935--1948, Sept 2015.

\bibitem{scin}
M.~Niu, J.~Cheng, and J.~F. Holzman, ``Error rate performance comparison of
  coherent and subcarrier intensity modulated optical wireless
  communications,'' \emph{J. Opt. Commun. Netw.}, vol.~5, no.~6, pp. 554--564,
  Jun 2013.

\bibitem{62}
A.~Prudnikov and Y.~A. Brychkov, \emph{INTEGRAL AND SERIES, Volume 3, More
  Special Functions}, Computing Center of the USSR Academy of Sciences, Moscow,
  1990.

\bibitem{mod1}
E.~Zedini and M.~S. Alouini, ``Multihop relaying over im/dd fso systems with
  pointing errors,'' \emph{Journal of Lightwave Technology}, vol.~33, no.~23,
  pp. 5007--5015, Dec 2015.

\bibitem{68}
\BIBentryALTinterwordspacing
``The wolfram functions site.'' [Online]. Available:
  \url{http://functions.wolfram.com}
\BIBentrySTDinterwordspacing

\bibitem{63}
P.~K. Mittal and K.~C. Gupta, ``An integral involving generalized function of
  two variables,'' \emph{Proceedings of the Indian Academy of Sciences -
  Section A}, vol.~75, no.~3, pp. 117--123, 1972.

\bibitem{bfox}
A.~Soulimani, M.~Benjillali, H.~Chergui, and D.~B. da~Costa, ``Performance
  analysis of {M-QAM} multihop relaying over mmwave weibull fading channels,''
  \emph{CoRR}, vol. abs/1610.08535, 2016.

\bibitem{e1}
Y.~Zhao, R.~Adve, and T.~J. Lim, ``{Symbol error rate of selection
  amplify-and-forward relay systems},'' \emph{IEEE Communications Letters},
  vol.~10, no.~11, pp. 757--759, November 2006.

\bibitem{e4}
J.~Vazifehdan and J.~H. Weber, ``{Symbol error rate of space-time coded
  multi-antenna wireless cooperative networks},'' in \emph{VTC Spring 2008 -
  IEEE Vehicular Technology Conference}, May 2008, pp. 1453--1457.

\bibitem{65}
H.~Jakuszenkow, ``On properties of the generalized gamma distribution,''
  \emph{Demonstratio Mathematica}, vol.~7, no.~1, pp. 13--22, 1974.

\bibitem{67}
A.~Prudnikov, \emph{Integrals and Series: Volume 1: Elementary
  Functions}.\hskip 1em plus 0.5em minus 0.4em\relax Taylor \& Francis, 1986.

\end{thebibliography}
\end{document}